\documentclass[12pt]{article}
\usepackage[utf8]{inputenc}
\usepackage{amsmath,setspace,geometry}
\usepackage{amsthm}
\usepackage{amsfonts}
\usepackage[shortlabels]{enumitem}
\usepackage{rotating}
\usepackage{pdflscape}
\usepackage{graphicx}
\usepackage{bbm}
\usepackage[dvipsnames]{xcolor}
\usepackage{hyperref}
\hypersetup{colorlinks=true, linkcolor= BrickRed, citecolor = BrickRed, filecolor = BrickRed, urlcolor = BrickRed, hypertexnames = true}
\usepackage[]{natbib} 
\bibpunct[:]{(}{)}{,}{a}{}{,}
\usepackage[english]{babel}
\usepackage{float}
\usepackage{caption}
\usepackage{subcaption}
\usepackage{booktabs}
\usepackage{pdfpages}
\usepackage{threeparttable}
\usepackage{lscape}
\usepackage{bm}

\usepackage{multirow}
\usepackage{booktabs}
\usepackage{adjustbox}
\bibpunct[:]{(}{)}{,}{a}{}{,}
\newtheorem{hypothesis}{Hypothesis}
\usepackage{setspace}

\title{Reference Points, Risk-Taking Behavior, and Competitive Outcomes in Sequential Settings}
\author{Masaya Nishihata\thanks{\href{mailto:}{nishihata-masaya@keio.jp}, Graduate School of Economics, Keio University and Mitsubishi UFJ Research and Consulting Co., Ltd.}, Suguru Otani\thanks{\href{mailto:}{suguru.otani@e.u-tokyo.ac.jp}, Market Design Center, Department of Economics, The University of Tokyo}\thanks{We thank Yastora Watanabe for his valuable advice on the draft and for sharing his experience in official bench press competitions. We also thank Nobuyuki Hanaki, Hidehiko Ichimura, Takeshi Murooka, and Taisuke Otsu for their valuable advice and Kenji Utagawa for his excellent research assistance. We also thank participants at the 18th Association of Behavioral Economics and Finance Conference and the 19th Applied Econometrics Conference. This work was selected for the 18th Behavioral Economics Society Encouragement Award.
This work was supported by JST ERATO Grant Number JPMJER2301 and JST SPRING Grant Number JPMJSP2123, Japan.}}
\date{
First version: September 20, 2024\\
Current version: \today
}

\begin{document}

\maketitle

\begin{abstract}
    Understanding how competitive pressure affects risk-taking is crucial in sequential decision-making under uncertainty. This study examines these effects using bench press competition data, where individuals make risk-based choices under pressure. We estimate the impact of pressure on weight selection and success probability. Pressure from rivals increases attempted weights on average, but responses vary by gender, experience, and rivalry history. Counterfactual simulations show that removing pressure leads many lifters to select lower weights and achieve lower success rates, though some benefit. The results reveal substantial heterogeneity in how competition shapes both risk-taking and performance.
\end{abstract}

\newpage
\section{Introduction}

Understanding the behavioral effects of reference points, particularly in the context of pressure and risk-taking, is a critical area of research. The core concept is that an individual's assessment of an outcome is influenced not only by the outcome itself but also by how it compares to a reference point \citep{tversky1992advances}. A significant challenge in this field is disentangling the effects of pressure associated with reference points on risk-taking behavior and outcomes. Unlike controlled experimental settings as in \cite{schwerter2024social}, where risk-taking can be modeled as a binary lottery, field data present complexities that make such decompositions difficult. In this paper, we utilize a sequential competition setting and panel dataset from official bench press competitions to explore the interplay between competitive pressure arising from reference points and risk-taking behavior. This dataset enables us to observe weight attempts, outcomes, rankings, and the exogenous pressure exerted by rivals over time. \textcolor{black}{In particular, we treat a lifter's interim ranking during the competition as the relevant reference point, which naturally emerges from the sequential structure of attempts. This implies that potential changes in rank are experienced as gains or losses, thereby generating competitive pressure.} By examining how pressure influences subsequent risk-taking decisions, we aim to enhance the understanding of behavior under uncertainty and contribute to discussions on optimizing competition design to maximize performance.

The primary contribution of this study is the decomposition of the effects of pressure arising from reference points into distinct elements: the choice of weight attempt (analogous to a lottery choice) and the probability of success (analogous to a lottery gain). \textcolor{black}{In this setting, choosing a heavier weight corresponds to selecting a lottery with a lower probability of success but a higher potential payoff in terms of achieved ranking, whereas choosing a lighter weight implies the opposite trade-off. We therefore interpret risk-taking as the selection of heavier attempt weights conditional on a lifter's ability.} Against this backdrop, the structure of official bench press competitions provides an ideal context to study the relationship between reference points and risk-taking within the sequential game setting of weight declaration and lifting stages, where exogenous pressure is present. Whether the lift is successful depends not only on the chosen weight but also on the competitive pressure arising from the possibility that the lifter's relative position may change during the competition. We also explore alternative competition designs that manipulate the availability of pressure-related information. Through this investigation, we seek to answer an analogous question in real-world competitive settings: What would be the impact on risk-taking behavior and expected outcomes if the pressure exerted by rivals were removed?

Our results demonstrate the role of pressure in shaping lifters' behavior during competition. Pressure from both lower- and higher-ranked rivals leads to increased attempt weights across both the second and third attempts, reflecting lifters' tendency to challenge themselves under competition. 

Regarding success probability, lifters under pressure from below show a slightly higher likelihood of success, possibly due to the motivation to defend their rank. However, when attempting to surpass higher-ranked rivals, success rates consistently decline, highlighting the difficulty of overtaking stronger competitors under pressure. This pattern is most pronounced in the third attempt, where fatigue and psychological strain may amplify the challenge. These results suggest that while competitive pressure can encourage bold strategies, it also introduces constraints that limit execution success in decisive moments.

Heterogeneity analyses further reveal that responses to pressure vary by gender, experience, and rivalry history. Male lifters exhibit stronger responses to pressure from both lower- and higher-ranked rivals in attempt selection, whereas female lifters show a more muted response, especially under upward pressure. Female lifters experience a sharper decline in success probability when attempting to overtake a higher-ranked rival, suggesting a heightened sensitivity to stress. More experienced lifters are more sensitive to higher-ranked competitors, reinforcing the idea that competitive awareness strengthens with experience. Additionally, lifters with a history of repeated rivalry encounters show heightened sensitivity to pressure, suggesting that familiarity enhances their ability to navigate competition. For success probabilities, while pressure from lower-ranked rivals can improve outcomes, particularly among experienced lifters, upward pressure generally reduces them, with stronger effects among female competitors. Lifters with strong rivalry history show a small negative impact from upward pressure but a large positive impact from lower pressure. These findings confirm substantial heterogeneity.

To examine how pressure shapes the balance between ambition and feasibility in decision-making, we simulate counterfactual competition designs in which such pressure is removed. The results reveal consistent directional patterns—more conservative attempts—but also considerable heterogeneity in expected outcomes. While a subset of lifters benefit from the absence of pressure, others experience lower success rates and reduced expected outcomes. These differences reflect the uneven role of external pressure: for some, it enables focus and composure; for others, it disrupts execution or leads to overly cautious choices.

More broadly, in decision-making contexts involving sequential risky choices, individuals may experience diverging outcomes depending on how they respond to the presence or absence of external competition. Particularly in the later stages of competition, ignoring external pressure and focusing on personal strategy may enhance outcomes for some, while others rely on competitive cues to sustain performance. These findings reinforce the idea that responses to pressure are highly individual, suggesting that the effectiveness of competition design depends on how one reacts to competition.

\subsection{Related literature}

This paper contributes to three strands of the literature: pressure from reference points, risk behavior under pressure, and competition design.

First, our paper contributes to the growing body of literature on reference points in sports competition. We follow the framework of \cite{o2018reference}, who suggest a variety of reference points, including the status quo (e.g., prior wealth), past experience, focal outcomes, aspirations, expectations, and other possibilities such as norms and social comparisons. However, the determination of reference points is often arbitrary and context-dependent \citep{dellavigna2018structural}, with these effects often combined in actual sports settings.\footnote{\cite{baillon2020searching} use Bayesian hierarchical modeling to estimate the marginal posterior distributions of six reference point rules. They found that subjects most frequently used the status quo and MaxMin as reference points.} Table \ref{tab:list_of_papers} summarizes the data, reference points, classifications, and the loss-gain domain of the utility function in related studies, although the classification is not perfect.

Salient numbers, such as round finish times in marathons \citep{allen2017reference,soetevent2022short} and predetermined par scores in golf \citep{pope2011tiger}, are well-known as focal reference points. Runners adjust their effort levels to meet target times, and golfers are significantly less accurate when attempting shots ahead of a predetermined reference point, labeled as ``par.'' \footnote{Beyond sports competition, \cite{rees2018quantifying} provides evidence implying that individuals have reference-dependent preferences, with zero tax due serving as the reference point. As experimental evidence, \cite{abeler2011reference} conducted an experiment manipulating the fixed pay amount as a reference point in a real-effort task. In this experiment, either fixed pay or piecework pay for a correct answer was randomly determined after task completion. They found that subjects were more likely to stop when the expected piecework pay equaled the fixed pay, exerting more effort when the fixed pay was high. \cite{corgnet2015goal} also show that goal setting by managers is most effective for team workers when monetary incentives are strong.} Goals set by athletes based on past or current performances are treated as reference points in the form of status quo, past experience, and expectations. For example, self-reported target times in marathons \citep{markle2018goals} and personal bests in chess \citep{anderson2018personal,gonzalez2024personal} serve as natural reference points, whether or not they function explicitly as goals. Theoretically, \cite{heath1999goals} and \cite{koch2016goals} discuss the concept of goals as reference points. In our data on bench press competition, we do not observe prevalent bunching around round weights, likely because the categories in which lifters compete are narrowly defined by characteristics such as age, body weight, and gender, so that competition is localized within relatively small groups and ranking differences are often determined by small weight margins. Additionally, unlike professional sports that provide substantial rewards, bench press competitions are amateur events with no direct monetary incentives. This setting highlights lifters' self-motivation to surpass personal bests and rank updates based on recent outcomes, independent of financial incentives. The current rank is the reference point, and each lifter is expected to have stronger incentives to avoid a large loss from a rank decrease than to pursue a relatively small gain from a rank increase.

In terms of pressure from rivals' outcomes, the phenomenon of ``choking under pressure''---where performance declines under stress---has been extensively studied. For instance, large-sample studies reject the hypothesis that first-mover advantage in soccer or being slightly behind increases the likelihood of winning in basketball, football, or rugby \citep{kocher2012psychological,teeselink2023does}.\footnote{Similar studies include the effects of spectators on penalty shootouts in soccer \citep{dohmen2008professionals}, free throws and playoff performance in basketball \citep{hickman2015impact,boheim2019choking,cao2011performance,toma2017missed,morgulev2018choking}, the impact of bookmakers' odds on soccer play \citep{bartling2015expectations}, putting performance in golf \citep{hickman2015impact}, the influence of decisive moments on dart throws \citep{teeselink2020incentives}, and biathlon performance \citep{harb2019choking,lindner2017choking}, the effect of previous game outcomes on current outcomes in hockey \citep{kniffin2014within}, and gender differences in pressure in tennis \citep{cohen2017choking,paserman2023gender}.} While some papers implicitly link the cause of pressure to reference points, we do not explicitly classify pressure under the loss-gain utility framework, as the specific cases do not fit neatly within that domain. Our study also relates to peer effects in terms of externalities from rivals' abilities and performance. \cite{guryan2009peer}, using random assignment in high-stakes golf tournaments, find no evidence of peer effects on average. In contrast, \cite{yamane2015peer} observe positive peer effects in swimming competitions, particularly suggesting that being chased by peers with lower personal bests enhances performance. These mixed results indicate that peer effects vary across different contexts, with \cite{nishihata2022heterogeneous} showing that peer effects among speed skaters depend on race distance and prize money. In our study, we leverage the sequential timing of observing rivals' declarations and outcomes, declaring the attempt weight, and executing the lift to decompose the effects of rivals' declarations and outcomes on both weight selection and the success probability of the lift.

Second, this paper contributes to the literature on the effect of reference points on risk-taking actions. The most closely related studies are \cite{puente2019reference} and \cite{genakos2012interim}. \cite{puente2019reference} examines the effect of current scores on the tendency to take high-risk shots in tennis. The author follows \cite{ely2017agents}, using the approach developed by \cite{klaassen2009efficiency} (a structural model) to show that first serves should be riskier than second serves, and using two first serves (or two second serves) is suboptimal. \cite{genakos2012interim} study the effects of interim ranking itself on weight selection and success probability using data from powerlifting competitions in the Olympic Games and World and European Championships between 1990 and 2006. In the experimental economics literature, \cite{schwerter2024social} conducted experiments on risk-taking behavior when subjects were exposed to exogenously predetermined peer earnings. The author found that subjects exposed to higher peer earnings increased their risk-taking behavior. Similarly, \cite{post2008deal} highlight the relevance of risk-taking behavior in path-dependent contexts, using data from the TV show ``Deal or No Deal'' and a replicated experiment where subjects choose to either ``deal'' for a sure prize or ``no deal'' to continue for an uncertain prize. Our setting is better suited than these examples for several reasons. First, weight selection conditional on personal bests and rivals' outcomes is analogous to choosing a lottery, where the probability of successfully lifting the weight can be inferred from data by both lifters and econometricians. Second, lifting weights involves a simple, monoarticular action, which leads to relatively fewer external factors influencing outcomes compared with other sports, including powerlifting \citep{genakos2012interim}. Third, unlike \cite{genakos2012interim}, we exploit the sequential competition design, which explicitly defines the information revealed to each lifter. Fourth, we use a universe of official bench press competitions with a richer sample size and set of variables, which have recently been recognized as important for detecting behavioral effects in sports data \citep{teeselink2023does}.

Third, our paper contributes to the literature on competition design aimed at maximizing each lifter's score within a group. This contribution parallels general frameworks in competition design within organizations and is related to contest theory, although sabotage and uncooperative behavior, which can occur in real-world workplace competition, are less likely in sports competition.\footnote{\cite{drago1998incentives} find that strong promotion incentives reduce helping behavior, even though individual effort increases. \cite{bandiera2005social} show that fruit pickers reduce their effort under relative incentives when their peers are friends, due to the negative externality of individual effort on others' payoffs.} As outlined by \cite{lazear2013personnel}, relative performance evaluation (RPE) exists in two forms. The first allows firms to filter out common shocks by comparing an individual's performance to that of a peer group. \cite{gibbons1990relative} provide empirical evidence suggesting that RPE was used in the compensation contracts of U.S. CEOs from 1974 to 1986. The second form of RPE involves fixed rewards based on the performance rankings of participants, commonly referred to as a rank-order tournament. In tournament theory, as modeled by \cite{lazear1981rank}, large differences in payoffs across ranks motivate those at lower levels to exert more effort, while some risk-averse agents may exert less effort when the importance of winning is diminished. Rank-order tournaments have a wide range of applications, including sports competitions. For example, consistent with tournament theory, \cite{brown2011quitters} shows that the presence of a superstar highly likely to win reduces the performance of other participants. In our context, the unique structure of official bench press competitions allows us to compare alternative competition settings to the current format through simulations based on our estimates.

Following the seminal model by \cite{lazear1981rank}, tournament theory has been extended in various directions to better reflect behavioral patterns and decision-making processes observed in real-world settings. Two particularly relevant strands of this literature focus on incorporating risk-taking decisions instead of traditional effort choices and integrating reference-dependent preferences into contest frameworks. \cite{hvide2002tournament}, \cite{hvide2003risk}, and \cite{taylor2003risk} incorporate risk-taking behavior into tournament models, analyzing how participants' strategic risk choices can be influenced by their relative positions and incentives within the tournament structure. \cite{nieken2010risk} show that the correlation of risk outcomes affects how both leading and trailing participants adjust their risk strategies. \cite{gill2010fairness} and \cite{dato2018expectation} extend tournament models by incorporating reference-dependent preferences, highlighting how loss aversion and outcome expectations can shape competitive behavior and effort provision. These extensions are closely related to our study, as they emphasize the importance of risk-taking behavior and reference-dependent preferences in tournament settings.


\begin{table}[ht]
    \begin{center}
    \caption{Empirical Sports and Experimental Papers Studying Reference Points}
    \label{tab:list_of_papers}
    \adjustbox{max width=\textwidth}{
    \begin{tabular}{|l|l|l|l|l|l|l|}
        \hline
        Paper & Data & Reference Point & Classification & Loss-Gain Domain \\
        \hline
        \cite{pope2011tiger} & Golf & Par For Each Hole & Focal Outcomes & Score For Each Hole \\
        \cite{pope2011round} & Basketball & Round Numbers In Performance & Focal Outcomes & Current Outcome  \\
        \cite{allen2017reference} & Marathon & Round Finish Time & Focal Outcomes & Finish Time \\
        \cite{essl2017choking} & Lab Experiment & Points (ECU) When $<,>$ 10sec & Focal Outcomes & Points From Real-Effort Task \\
        \cite{soetevent2022short} & Running Event & Round Finishing Time & Focal Outcomes & Finishing Time  \\
        \cite{schwerter2024social} & Lab Experiment & Peer Earnings & Focal Outcomes & Earnings From Binary Lotteries \\
        \cite{post2008deal} & TV Shows & Earlier Expectations & Expectations & Bank Offer \\
        \cite{abeler2011reference} & Lab Experiment & Amount Of Fixed Pay & Expectations & Piece Rate For Correct Answers \\
        \cite{bartling2015expectations} & Soccer & Expected Outcome By Odds & Expectations & Current Outcome \\
        \cite{markle2018goals} & Marathon & Self-Reported Target Time & Status Quo & Finish Time \\
        \cite{kniffin2014within} & Hockey & The Outcomes For Game 1 & Past Experience & Score In Game 2 \\
        \cite{anderson2018personal} & Chess Online & Personal Best Rating & Past Experience & Rating  \\
        \cite{gonzalez2024personal} & Chess In-Person & Personal Best Rating & Past Experience & Rating  \\
        \cite{berger2011can} & Basketball & Performance Of The Opponent & Social Comparison & Current Outcome \\  
        \cite{puente2019reference} & Tennis & Rival's Score & Social Comparison & Score \\
        \cite{teeselink2023does} & Four Sports & Performance Of The Opponent & Social Comparison & Current Outcome \\
        
        This Paper & Bench Press & Current Own Ranking & Social Comparison & Ranking \\
        \cite{baillon2020searching} & Lab Experiment & 6 Rules & Mixed & Lottery Options \\
        \hline
    \end{tabular}
    }
    \end{center}\footnotesize 
    \textit{Notes}: We follow the referent list of \cite{o2018reference}, which suggests a number of possibilities: the status quo (e.g., prior wealth), past experience, focal outcomes, aspirations, expectations, and other candidates including norms and social comparisons, although reference point determination is in many ways arbitrary and context-dependent \citep{dellavigna2018structural}, and these effects are often combined in actual sports settings. \\
    6 rules (Status Quo, MaxMin, MinMax, X at Max P, Expected Value, Prospect Itself), ECU=Experimental Currency Unit.
\end{table}

\section{Data}

\subsection{Data source}

We use data from \cite{OpenPowerlifting}, a community service project that creates a permanent and open archive of world powerlifting data.\footnote{\url{https://gitlab.com/openpowerlifting.} Data accessed on July 1, 2023.} 
The OpenPowerlifting database includes a wide range of fields from official powerlifting competitions. It contains competition details (federation name, date and location, and level: local, national, or international), athlete information (name, age, weight class, and gender), lift results for squat, bench press, and deadlift (attempted weights and success/failure status), Wilks score and other relative strength metrics, and rankings and historical records within competitions.

In this paper, we focus on bench press competitions, which represent the most popular division and are less complex than Squat-Bench-Deadlift (SBD) composite competitions. 
We use all official attempt data for lifters aged 15-69 in competition categories based on age class, weight class, and gender, where the number of participants is at least two to capture competitive pressure.

\begin{table}[!htbp]
  \begin{center}
      \caption{Summary Statistics}
      \label{tb:summary_statistics} 
      
\begin{tabular}[t]{llrrrrr}
\toprule
Equipment &   & N & mean & sd & min & max\\
\midrule
Raw & Male & 175983 & 0.86 & 0.35 & 0.00 & 1.00\\
 & Personal best & 175983 & 124.39 & 58.77 & 0.00 & 310.00\\
 & First attempt weight & 175983 & 130.96 & 44.77 & 5.00 & 330.00\\
 & Second attempt weight & 175983 & 137.42 & 45.89 & 20.00 & 352.50\\
 & Third attempt weight & 175983 & 141.62 & 46.57 & 20.00 & 352.50\\
 & Successful first attempt & 175983 & 0.86 & 0.35 & 0.00 & 1.00\\
 & Successful second attempt & 175983 & 0.72 & 0.45 & 0.00 & 1.00\\
 & Successful third attempt & 175983 & 0.40 & 0.49 & 0.00 & 1.00\\
 & Best attempt & 169111 & 137.97 & 46.25 & 0.00 & 325.00\\
 & Age & 175983 & 31.77 & 11.22 & 13.00 & 69.00\\
 & Bodyweight & 175983 & 85.35 & 20.37 & 24.90 & 240.00\\
Single-ply & Male & 76681 & 0.80 & 0.40 & 0.00 & 1.00\\
 & Personal best & 76681 & 133.97 & 80.06 & 0.00 & 402.50\\
 & First attempt weight & 76681 & 156.16 & 64.67 & 5.00 & 445.00\\
 & Second attempt weight & 76681 & 162.82 & 65.51 & 20.00 & 445.00\\
 & Third attempt weight & 76681 & 168.05 & 66.65 & 20.00 & 445.00\\
 & Successful first attempt & 76681 & 0.77 & 0.42 & 0.00 & 1.00\\
 & Successful second attempt & 76681 & 0.65 & 0.48 & 0.00 & 1.00\\
 & Successful third attempt & 76681 & 0.40 & 0.49 & 0.00 & 1.00\\
 & Best attempt & 70756 & 160.21 & 64.68 & 0.00 & 445.00\\
 & Age & 76681 & 32.32 & 12.93 & 13.00 & 69.00\\
 & Bodyweight & 76681 & 84.89 & 23.19 & 30.50 & 245.00\\
Multi-ply & Male & 14678 & 0.97 & 0.18 & 0.00 & 1.00\\
 & Personal best & 14678 & 157.23 & 95.59 & 0.00 & 425.00\\
 & First attempt weight & 14678 & 197.10 & 63.60 & 25.00 & 480.00\\
 & Second attempt weight & 14678 & 205.88 & 64.64 & 30.00 & 500.50\\
 & Third attempt weight & 14678 & 212.12 & 65.72 & 30.00 & 520.00\\
 & Successful first attempt & 14678 & 0.70 & 0.46 & 0.00 & 1.00\\
 & Successful second attempt & 14678 & 0.57 & 0.49 & 0.00 & 1.00\\
 & Successful third attempt & 14678 & 0.35 & 0.48 & 0.00 & 1.00\\
 & Best attempt & 12864 & 201.43 & 63.12 & 0.00 & 430.00\\
 & Age & 14678 & 32.20 & 10.10 & 13.00 & 69.00\\
 & Bodyweight & 14678 & 94.69 & 19.65 & 34.40 & 207.80\\
\bottomrule
\end{tabular}

  \end{center}\footnotesize
  \textit{Sources}: The OpenPowerlifting database. 
\end{table} 

Table \ref{tb:summary_statistics} presents summary statistics across three types of powerlifting equipment: raw, single-ply, and multi-ply. Each category includes attributes such as personal best, first, second, and third attempt weights, successful attempts, best attempt, age, and body weight. The raw category has the largest number of participants, with 175,983 individuals, an average personal best of 124.39 kg, and a mean body weight of 85.35 kg. The single-ply category includes 76,681 participants, with an average personal best of 133.97 kg and a mean body weight of 84.89 kg. In the multi-ply category, there are 14,678 participants, with an average personal best of 157.23 kg and a mean body weight of 94.69 kg.

Across all equipment types, the mean weights lifted increase from the first to the third attempts, and the percentage of successful attempts decreases with each successive attempt. This indicates that while lifters attempt heavier weights in successive attempts, the success rate declines. Approximately 86\% of participants in the raw category, 80\% in the single-ply category, and 97\% in the multi-ply category are male. The average age of participants is around 31.77 years for raw lifters, 32.32 years for single-ply lifters, and 32.20 years for multi-ply lifters. These statistics highlight key performance trends and demographic characteristics in competitive powerlifting, emphasizing differences in performance and success rates among different equipment types and attempts.

Personal best weights of lifters who have never participated in an official bench press competition are recorded as zero. In the empirical analysis, we focus on the second and third attempts and treat the realized outcome in the first attempt as the personal best for such lifters because the success probability in the first attempt is 86\% for raw lifters, 77\% for single-ply lifters, and 70\% for multi-ply lifters.\footnote{\textcolor{black}{Appendix \ref{sec:first_attempt_rival_pb} shows the supplemental analysis of the first attempts, which shows similar findings to our main results on the second and third attempts.}}

\subsection{Game setting}
The objective of each lifter is to lift the maximum weight possible from a prone position on a flat bench to improve their own rank.
In case of a tie (same weight lifted), the lighter lifter or the lifter who achieved the weight first (depending on the federation's rules) is ranked higher.
Competitors are divided into categories based on weight class, age class, equipment class, and gender to ensure fair competition.\footnote{See details in the International Powerlifting Federation (IPF) Technical Rules Book.}

\paragraph{Competition structure}
Competitors declare their opening weight (first attempt) before the competition starts. 
The declared weights determine the initial lifting order.
Lifters are arranged in ascending order based on their declared opening attempts.
That is, the lifter with the lightest weight goes first, followed by the next lightest, and so on.

Each lifter has three attempts.\footnote{
Lifting Procedure for each lifter at each attempt is officially divided into the following.
\begin{enumerate}
    \item Setup: The lifter lies on their back on the bench, with feet flat on the floor or on the bench's footrests, ensuring contact with the bench throughout the lift.
    \item Grip: The lifter grips the barbell, usually slightly wider than shoulder-width apart.
    \item Unracking: With the help of spotters if needed, the lifter unracks the barbell and holds it with arms fully extended above their chest.
    \item Lowering: The barbell is lowered in a controlled manner to touch the chest, ensuring it is stable and paused for a moment.
    \item Press Command: The judge gives a “Press" command once the barbell is motionless on the chest. The lifter then presses the barbell back to the starting position with arms fully extended.
    \item Rack Command: Upon completion, and with the barbell in a stable position, the judge gives the “Rack" command, signaling the lifter to return the barbell to the rack.
\end{enumerate}}
Each attempt is judged based on control, stability, and completion of lifting.
A lift is considered successful if the lifter adheres to all rules and completes the lift according to the judge's commands, as determined by at least two of the three referees. 
After completing an attempt, the lifter must declare the weight for their next attempt within a specific time frame (usually 1 minute after their attempt).
Once a weight is declared, it generally cannot be decreased, only increased.\footnote{\textcolor{black}{Lifters therefore tend to begin with relatively conservative weights and choose heavier weights in later attempts. This pattern can be interpreted as a natural strategic response to the need to avoid finishing with no recorded result, implying greater risk-taking in later rounds.}}
The order of lifters is adjusted after each round of attempts based on the declared weights for the next attempt, again proceeding from lightest to heaviest. This ensures that the competition is efficient and fair.

Concretely, in the first attempt, all lifters complete their first attempt in the sequential order of the lightest to heaviest declared weights.
In the second attempt, the order is recalculated based on the declared weights for the second attempt. Lifters who declared lighter weights for their second attempt will lift first.
In the third attempt, the order is again recalculated for the third attempt based on the new declared weights.
If a lifter fails an attempt, they can choose to reattempt the same weight or increase it for their next attempt. The order will still be based on the declared weight for the next attempt, regardless of whether the previous attempt was successful.

Each lifter's highest successful attempt is recorded as their final score.
If two lifters achieve the same highest lift, the lighter lifter or the lifter who achieved the weight first (depending on the federation's rules) ranks higher.

\paragraph{Example of attempt order:}

\begin{figure}[!ht]
\begin{center}
\includegraphics[height = 0.4\textheight]{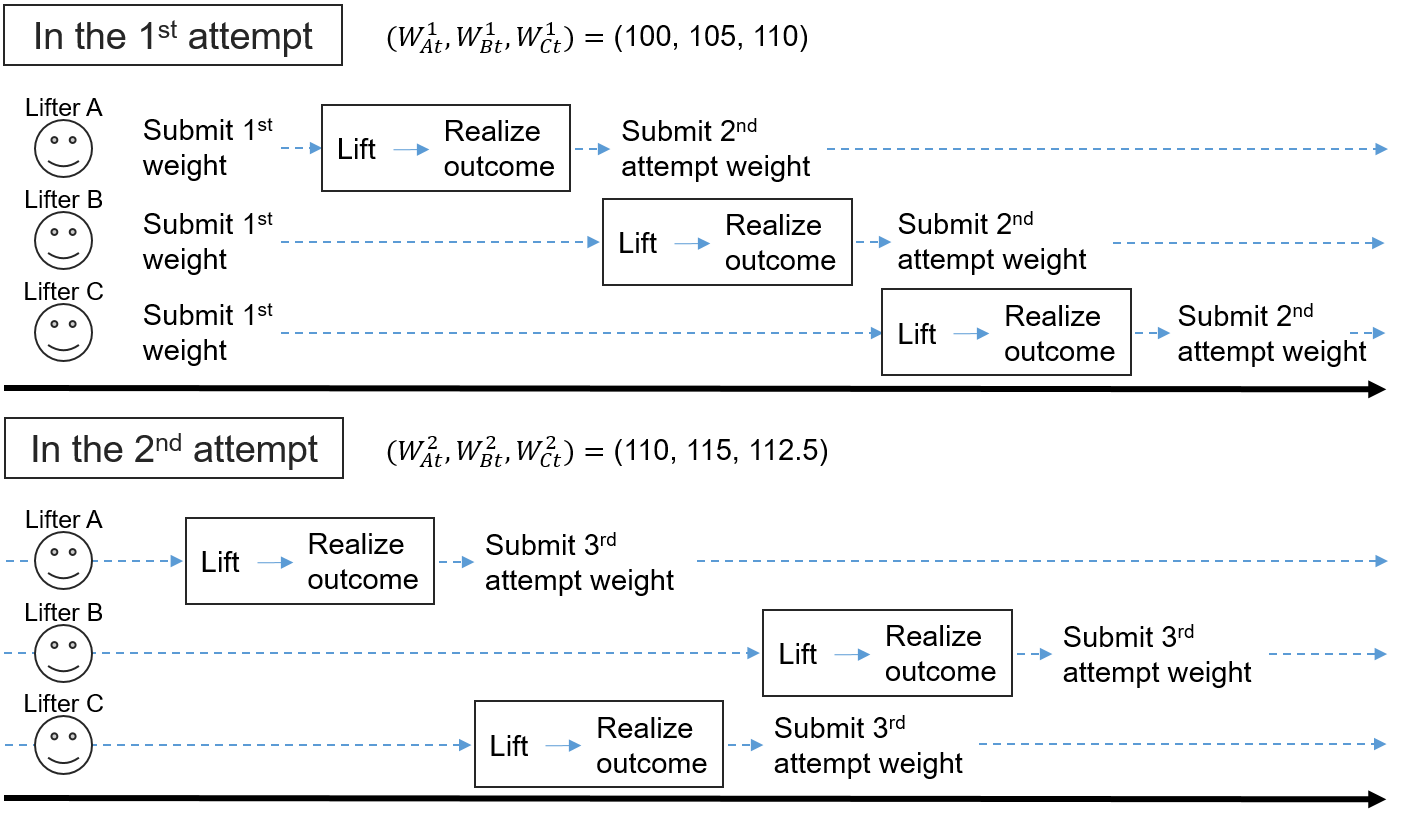}
\end{center}
\caption{The Structure of the First and Second Attempts: Three Lifters Case}\footnotesize
\textit{Notes}: The box indicates that the lifter appears on the stage. $W_{it}^{j}$ is lifter $i$'s attempt weight in the $j$-th attempt in competition $t$.
\label{fg:weightlifting_figure}
\end{figure}

To highlight the information available to each lifter at each attempt, consider a competition with three lifters (A, B, and C). 
The timeline of the first and second attempts is shown in Figure \ref{fg:weightlifting_figure}.
For the opening attempts, lifters $A$, $B$, and $C$ declare 100 kg, 105 kg, and 110 kg before the competition. 
The order is $A$, $B$, and $C$.
Lifter $A$ attempts 100 kg, and the outcome is realized.
Next, lifter $B$ attempts 105 kg.
At the same time, lifter $A$ declares the next attempt weight (110 kg) within one minute, conditional on $A$'s own first outcome and on $B$'s and $C$'s attempt weights, before $B$'s outcome is realized because lifter $B$ is attempting 105 kg at that time.
Lifter $B$'s outcome is realized.
Next, lifter $C$ attempts 110 kg.
At the same time, lifter $B$ declares the next attempt weight (115 kg) within one minute, knowing $B$'s own first outcome, $A$'s outcome and next attempt, and $C$'s attempt weight, before $C$'s outcome is realized because lifter $C$ is attempting 110 kg at that time.
Lifter $C$'s outcome is realized.
Lifter $C$ declares the next attempt weight (112.5 kg) within one minute, knowing $A$'s and $B$'s outcomes, their next attempts, and $C$'s own first outcome.
For the second attempts, the new order is $A$ (110 kg), $C$ (112.5 kg), and $B$ (115 kg).
After the same procedure, the order for the third attempt is similarly determined by the declared weights in the second attempt.

\subsection{\textcolor{black}{Reference point in bench press competition}}\label{sec:reference_point}

A wide range of reference points have been discussed in the literature, including the status quo, past experience, focal outcomes, aspirations, expectations, and norms \citep{o2018reference}. In bench press competitions, potential candidates include personal best records, salient numbers, and opponents' outcomes. In this study, however, we focus on a lifter's current ranking within the competition as the reference point. Under this definition, competitive outcomes are naturally evaluated in terms of whether a lifter improves upon, maintains, or loses their current rank as the competition unfolds.

Bench press competitions are explicitly rank-based. Athletes compete to achieve a higher placement, and rankings are determined by the outcomes of a finite sequence of attempts. Because each competition consists of three attempts, intermediate rankings become observable from the second attempt onward. This allows lifters to assess how their relative position may change in response to rivals' declared weights and realized outcomes, making current ranking a particularly salient benchmark for decision-making during the competition.

The sequential structure of the competition plays a crucial role in this setting. As discussed in Section \ref{sec:exogenous_pressure}, the timing of information revelation regarding rivals' attempts and outcomes generates exogenous variation in competitive pressure relative to the current ranking. While individual lifters may differ in the extent to which they attend to various potential reference points, focusing on current ranking provides a natural and empirically tractable way to study how pressure arising from relative position affects risk-taking behavior and performance in a sequential competition.

\subsection{Exogenous pressure}\label{sec:exogenous_pressure}

The above sequential game structure, in which each lifter does not know subsequent attempts and outcomes, provides exogenous variation in both attempt-weight choice and lifting success, analogous to lottery choice and realized payoff. 
We focus on the moves of the closest rivals to each lifter, namely the immediately lower- and higher-ranked rivals, which are critical for determining the final rank in the competition.

\paragraph{The choice of attempt weight}
For second-attempt weight choice, each lifter observes the first-attempt outcome and the second-attempt weight of the immediately lower-ranked rival. These are uncorrelated with the lifter's unobservable characteristics, such as body condition on the competition date, because the rival declares the second-attempt weight before observing the outcome of the focal lifter's second attempt, which---if observable---could serve as a proxy for the focal lifter's condition. The exogenous pressure affects the risk of being overtaken by the lower-ranked rival, that is, the risk of loss. 
Second, each lifter forms an expectation about the second-attempt weight of the immediately higher-ranked rival, although this value is not yet observed. We assume that each lifter can predict a higher-ranked rival's action based on information about the rival's past attempts and a rich set of covariates.\footnote{The assumption is supported by our data in Section \ref{sec:expected_rival_attempt_weight}. Strategic reasoning in intermediate weight selections is also inherently limited, as each subsequent attempt must exceed the current attempt. While incorporating dynamic and strategic considerations is theoretically appealing, doing so introduces significant complexity and falls beyond the scope of our empirical analysis.} Because these predictions are based solely on observable characteristics, they are independent of the lifter's own unobserved factors, such as temporary performance conditions. Because the first-attempt weight of the higher-ranked rival is heavier than the focal lifter's current best weight, overtaking that rival is more difficult. The exogenous pressure affects the challenge of overtaking the higher-ranked rival, that is, the risk of gain.

\paragraph{The success of lifting the attempt weight}
During the second-attempt lifting stage, each lifter knows the first- and second-attempt outcomes of the immediately lower-ranked rival. These are uncorrelated with the lifter’s unobserved characteristics, such as body condition on the competition date, because the lower-ranked rival completes both attempts without observing the focal lifter's second outcome. The exogenous pressure affects the risk of being overtaken by the lower-ranked rival, that is, the risk of loss. If the focal lifter fails the second attempt, the rival overtakes him. 
Second, each lifter also knows the second-attempt weight of the immediately higher-ranked rival. This declared weight is uncorrelated with the lifter's unobserved characteristics because, at the time of declaration, the higher-ranked rival has not yet observed the focal lifter's second outcome. If the focal lifter succeeds in the second attempt, he has a greater chance to overtake the rival. The exogenous pressure therefore affects the challenge of overtaking the higher-ranked rival, that is, the risk of gain. Note that this pressure matters only at the lifting stage.

\subsection{Data pattern}
We illustrate data patterns in attempt-weight choice and the success probability of an attempt conditional on the exogenous pressure defined above.
For illustration, we show the most popular competition category (Male, Raw, 24-39 age class). Similar patterns appear in other competition categories. We omit the first-attempt patterns because the success probability at that stage is above 75\%.

\paragraph{Attempt weight, success probability, and pressure}

\begin{figure}[!ht]
\begin{center}
\includegraphics[height = 0.40\textheight]{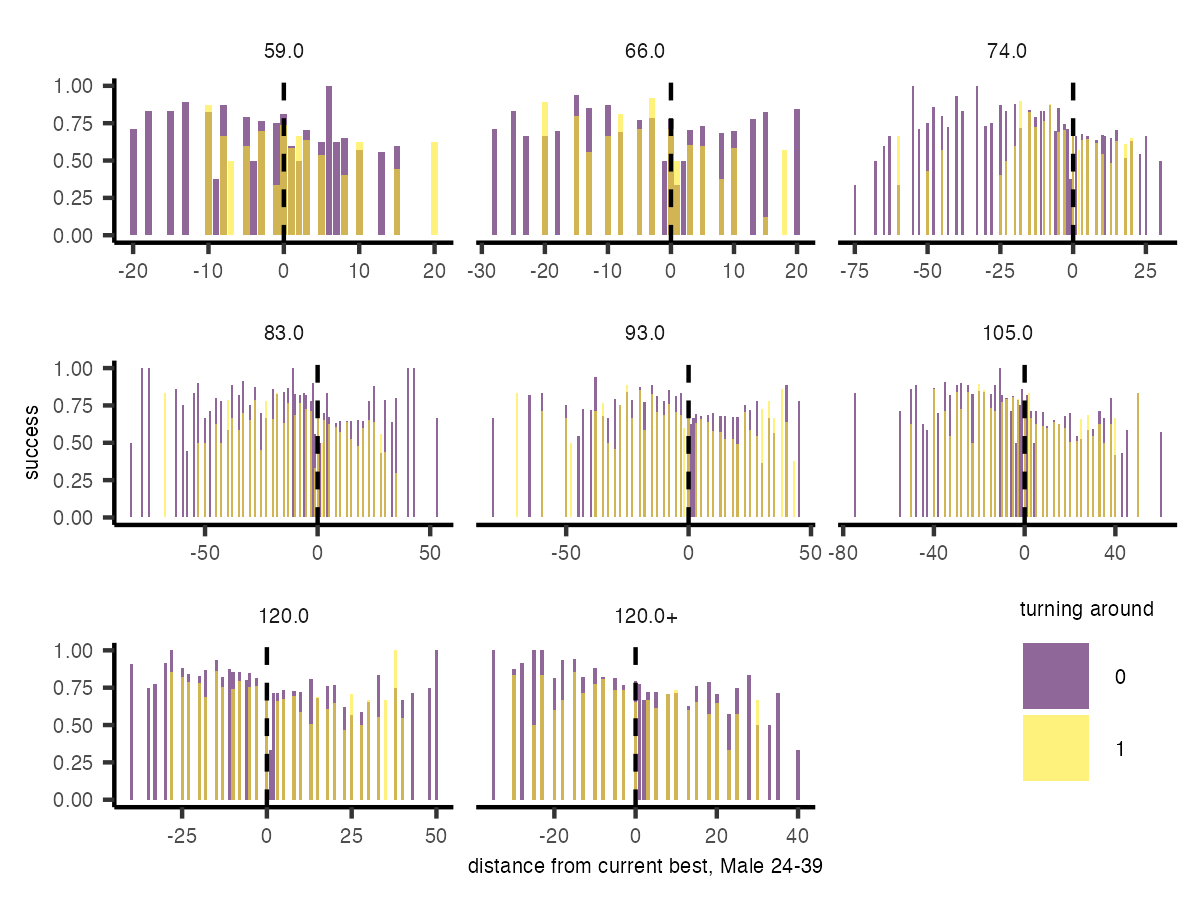}
\includegraphics[height = 0.40\textheight]{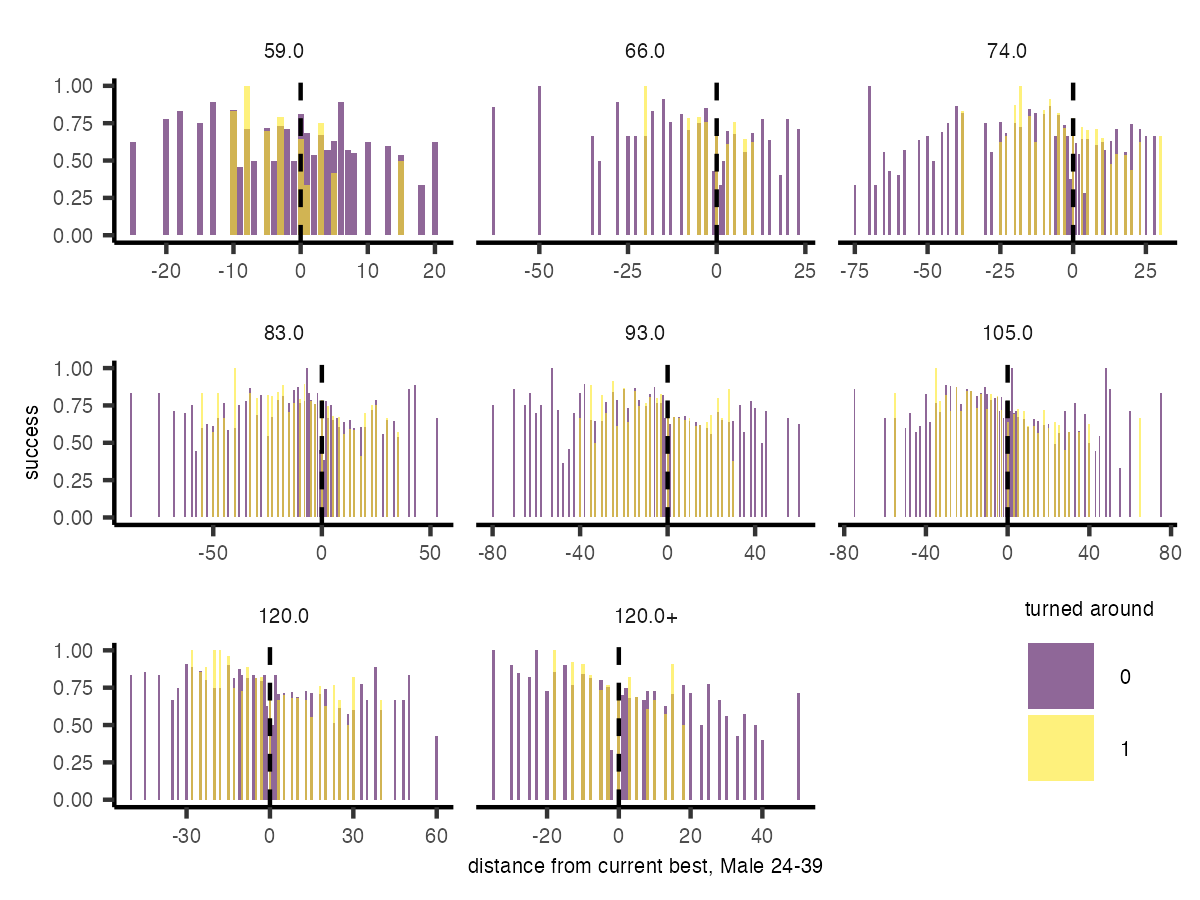}
\end{center}
\caption{Success Rate Conditional on Pressures at Second Attempt (Male, Raw, 24-39 age class)}\footnotesize
\textit{Notes}: The distance from the current best was rounded to the nearest integer. Distance categories with five or fewer lifters were excluded from the plot to avoid skewing the success probabilities to 0 or 1.
\label{fg:success_rate_Raw24-39M_apply_bench2kg_turning_around_one_higher_rank_player_bench1kg}
\end{figure}

Figure \ref{fg:success_rate_Raw24-39M_apply_bench2kg_turning_around_one_higher_rank_player_bench1kg} displays success probabilities for second bench press attempts across different weight classes, showing the highest success rates when the attempted weight is slightly below the lifter's personal best. Across almost all classes, success peaks around a distance of zero, with success probabilities typically ranging from 0.75 to 1.00 at this point. 
As the attempted weight deviates from the best, success rates drop significantly, often falling below 0.25 at more extreme deviations, particularly for heavier attempts. 
In heavier weight classes (120 kg and 120+ kg), the range of successful attempts is slightly broader, but the general trend of declining success with larger deviations remains consistent across all weight categories.
While both pressured (``turning around'' or ``turned around'') and non-pressured groups follow similar patterns, the pressured group generally maintains slightly higher success probabilities, particularly near the lifters' personal best. However, variations exist across different weight categories, with some distances exhibiting no clear advantage for the pressured group.

\begin{figure}[!ht]
\begin{center}
\includegraphics[height = 0.40\textheight]{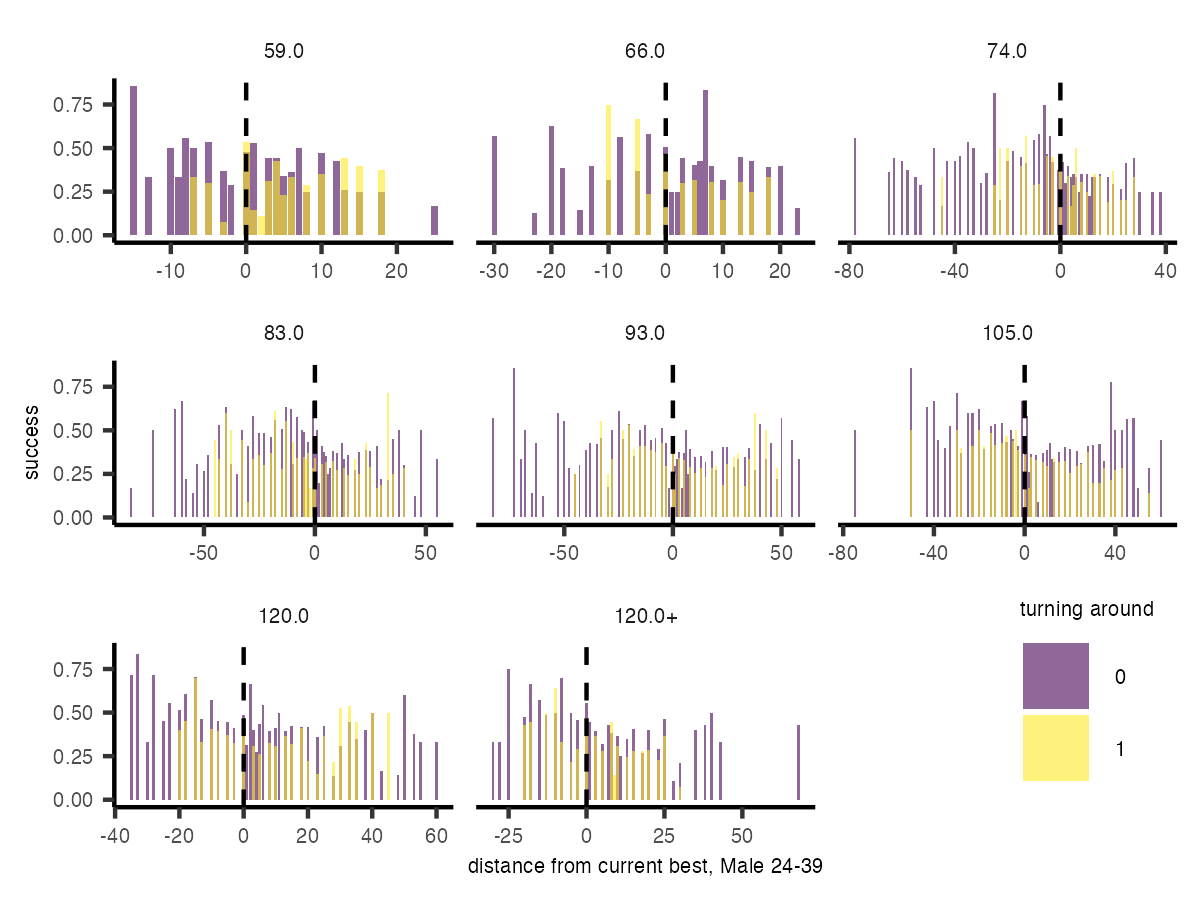}
\includegraphics[height = 0.40\textheight]{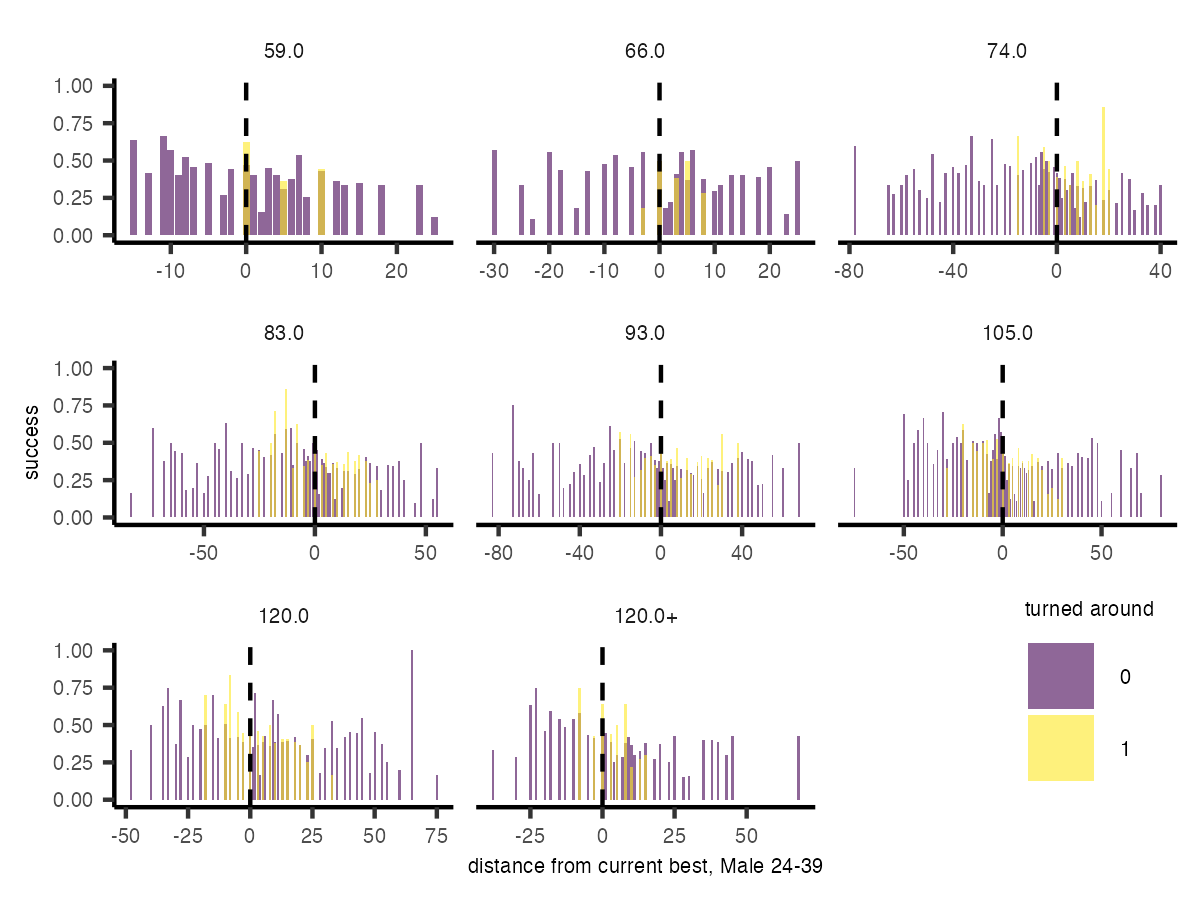}
\end{center}
\caption{Success Rate Conditional on Pressures at Third Attempt (Male, Raw, 24-39 age class)}\footnotesize
\textit{Notes}: The distance from the current best was rounded to the nearest integer. Distance categories with five or fewer lifters were excluded from the plot to avoid skewing the success probabilities to 0 or 1.
\label{fg:success_rate_Raw24-39M_apply_bench3kg_turning_around_one_higher_rank_player_bench2kg}
\end{figure}

Figure \ref{fg:success_rate_Raw24-39M_apply_bench3kg_turning_around_one_higher_rank_player_bench2kg} illustrates more distinct patterns for third bench press attempts. Compared to the second attempt, success rates in the third attempt are noticeably lower across all weight classes. While success rates during the second attempt peaked between 0.75 and 1.00 near the lifters' personal best, the third attempt generally shows lower peak success rates, ranging between 0.5 and 0.8. Additionally, the decline in success rates as lifters deviate from their personal best is more pronounced in the third attempt, with success probabilities frequently dropping below 0.25 at larger deviations. The distinction between the ``turning around'' groups (0 and 1) remains visible, though the overall lower success rates make the difference less pronounced in the third attempt. Overall, lifters appear to struggle more in the third attempt, particularly when attempting weights significantly different from their personal best, indicating a notable drop in success probability compared to the second attempt.

In summary, the data suggest that when lifters face pressure in critical situations—such as attempting to improve their rank—they tend to choose more challenging weights. However, unlike the second attempt, where pressured lifters often maintain higher success rates, the third attempt shows a more substantial decline in overall success probabilities, suggesting increased difficulty in securing successful lifts under pressure.

\paragraph{Who are the rivals?}

We examine the pressure exerted by rivals during competition. A natural question arises: who constitutes a rival for each lifter within a competition? Table \ref{tb:changes_in_ranking} illustrates the distribution of rank changes for lifters between their second and third attempts. The majority of lifters maintain their rank, as shown by the high proportion of zero rank changes (67\% in the second attempt and 72\% in the third attempt). A smaller percentage of lifters experience a rank increase ($+1$) between attempts (14\% in the second and 10\% in the third), while the proportion of those gaining two or more ranks is notably lower (5\% in the second and 3\% in the third). Similarly, lifters dropping one rank ($-1$) account for 8\% in the second attempt and 9\% in the third, with even fewer dropping two or more ranks (5\% and 6\%, respectively). Based on this distribution, we focus on the immediate competitors, specifically the lifter ranked directly below and the one ranked directly above, as these rivals most significantly influence rank changes.

\begin{table}[!htbp]
  \begin{center}
      \caption{Distribution of Rank Changes}
      \label{tb:changes_in_ranking} 
      
\begin{tabular}{ccc}
\toprule
Changes in ranking & 2nd & 3rd\\
\midrule
$\leq -2$ & 0.05 & 0.06\\
$-1$ & 0.08 & 0.09\\
$0$ & 0.67 & 0.72\\
$+1$ & 0.14 & 0.10\\
$+2 \leq$ & 0.05 & 0.03\\
\bottomrule
\end{tabular}

  \end{center}\footnotesize
  \textit{Notes}: We compute the distribution of rank changes using all achieved outcomes of all lifters in our data.
\end{table}

\section{Conceptual framework}
In this section, we develop a conceptual framework based on expected utility theory and prospect theory. Bench press competitions are organized in successive rounds, with interim rankings determined after each round. This setting naturally motivates a reference-dependent evaluation of competitive outcomes, where the interim rank serves as the reference point.

Following \cite{tversky1992advances}, we adopt a rank-dependent value function $v(\cdot)$ with reference point $r$ defined as
\begin{align*}
    v(x)=
    \begin{cases}
        (x-r)^{\alpha} & \text{if } x\ge r,\\
        -\lambda(r-x)^{\alpha} & \text{if } x<r,
    \end{cases}
\end{align*}
where $0<\alpha\le1$ and $\lambda>1$ capture diminishing sensitivity---concave curvature in the gain domain and convex curvature in the loss domain---and loss aversion, respectively. Here, $x$ denotes the rank of lifter $i$ at the end of a stage, and $r$ denotes lifter $i$'s interim rank prior to that stage. We adopt the convention that higher values correspond to better positions; accordingly, we reverse the ordering of the ordinal rank so that the best position takes the largest value. Under this convention, $x-r=+1$ corresponds to an improvement by one position and $x-r=-1$ to being overtaken by one position. For simplicity, we focus on local competition and assume $x-r\in\{-1,0,1\}$.\footnote{Table~\ref{tb:changes_in_ranking} shows that rank changes of two or more positions are uncommon (approximately 5\% of observations), so this restriction captures the empirically relevant margin. Moreover, when $x-r\in\{-1,0,1\}$, we have $|x-r|^{\alpha}=|x-r|$ for any $0<\alpha\le1$, implying that local asymmetry is governed by $\lambda$.} To maintain tractability and focus on empirically testable implications, we model the attempt and lifting stages separately and treat rivals' decisions as given from the focal lifter's perspective. \footnote{A fully dynamic equilibrium model with endogenous effort and strategic interaction would substantially complicate the framework and is beyond the scope of the present study.}

\subsection{Attempt stage}

At the attempt stage, lifter $i$ chooses an attempt weight $w_i$. We focus on local competition among three lifters: lifter $i$, the lower-ranked rival $L$ immediately below $i$, and the higher-ranked rival $H$ immediately above $i$ in the interim ranking at the start of the round. Let $y_i$ denote lifter $i$'s highest successfully lifted weight prior to the current round. Since attempt weights cannot be reduced once declared, the feasible set satisfies $w_i \ge y_i$. For any declared attempt weight $w$, let $q_i(w)\in[0,1]$ denote lifter $i$'s probability of success. We assume that heavier weights are more difficult to lift, so that $q_i'(w)<0$ for any $i$.

We first consider pressure from the lower-ranked rival. Suppose that the lower rival declares an attempt weight $w_L > y_i$. If lifter $i$ chooses $w_i$ such that $w_L > w_i$, lifter $i$ is overtaken if and only if the lower rival succeeds. The expected utility is then given by $EU \mid_{w_L > w_i} = -\lambda q_L(w_L) \le 0$. If instead lifter $i$ chooses $w_i > w_L$, overtaking occurs only when lifter $i$ fails and the lower rival succeeds. The expected utility becomes $EU \mid_{w_i > w_L} = -\lambda\bigl(1-q_i(w_i)\bigr)q_L(w_L) \le 0$.

Since $0 < 1-q_i(w_i) < 1$, we have
\begin{align*}
    EU \mid_{w_i > w_L} > EU \mid_{w_L > w_i}
    \quad \text{and} \quad
    \frac{\partial EU \mid_{w_i > w_L}}{\partial w_i}
    = \lambda q_L(w_L) q_i'(w_i) \le 0,
\end{align*}
which implies that lifter $i$ maximizes expected utility by choosing an attempt weight just above $w_L$.

When $w_L < y_i$, no rank change occurs regardless of the success or failure outcomes of either lifter. In this case, expected utility satisfies $EU(w_i) \mid_{y_i > w_L} = 0$, and lifter $i$ is indifferent over all feasible choices of $w_i$. \footnote{Note that this situation depends on the lower rival's choice and cannot be induced by lifter $i$'s own declaration.}

\begin{hypothesis}
    When the lower rival's declared attempt weight exceeds lifter $i$'s current best, an increase in competitive pressure from below---measured by $w_L - y_i$---induces lifter $i$ to choose a heavier attempt weight, i.e., to engage in greater risk-taking.
\end{hypothesis}

We next turn to pressure from the higher-ranked rival. The higher rival's declared attempt weight $w_H$ is not directly observed by lifter $i$ at the time of the attempt stage. However, as discussed in Section~\ref{sec:expected_rival_attempt_weight}, lifter $i$ can form expectations about $w_H$ based on observable characteristics.

If lifter $i$ chooses $w_i < y_H$, no rank change occurs regardless of the success or failure of $i$'s outcome. In this case, expected utility satisfies $EU \mid_{y_H > w_i} = 0$, and lifter $i$ is indifferent over all such choices of $w_i$.

By contrast, choosing $w_i > y_H$ creates a possibility of overtaking the higher-ranked rival. We therefore compare the expected utility from alternative choices with $w_i > y_H$, taking as given lifter $i$'s expectations about the higher rival's attempt weight $w_H$. Given an expectation of $w_H$, lifter $i$ may declare either $w_i^- \in (y_H,w_H)$ or $w_i^+ > w_H$. If $w_H > w_i^- > y_H$, lifter $i$ overtakes the higher rival only if lifter $i$ succeeds and the higher rival fails. Expected utility is therefore $EU \mid_{w_H > w_i^- > y_H} = q_i(w_i^-)\,\bigl(1-q_H(w_H)\bigr) \ge 0$. If $w_i^+ > w_H \ge y_H$, lifter $i$ overtakes the higher rival whenever lifter $i$ succeeds. Expected utility is then $EU \mid_{w_i^+ > w_H} = q_i(w_i^+) \ge 0$.

Since choosing $w_i<y_H$ yields zero expected utility while $w_i>y_H$ yields weakly positive expected utility, lifter $i$ prefers to select an attempt weight exceeding $y_H$. Among choices with $w_i > y_H$, expected utility is decreasing in $w_i$ in both relevant regions. In particular,
\begin{align*}
    \frac{\partial EU \mid_{w_i^+ > w_H}}{\partial w_i^+}
    = q_i'(w_i^+) < 0
    \quad \text{and} \quad
    \frac{\partial EU \mid_{w_H > w_i^-}}{\partial w_i^{-}}
    = q_i'(w_i^{-})\,\bigl(1-q_H(w_H)\bigr) \le 0.
\end{align*}
Therefore, the optimal choice lies at the boundary of each region: either just above $y_H$ or just above the expected $w_H$.

Comparing these two boundary choices, lifter $i$ prefers declaring a weight just above the expected $w_H$ rather than just above $y_H$ if and only if
\begin{align*}
    \frac{q_i(w_i^+)}{q_i(w_i^-)} > 1 - q_H(w_H).
\end{align*}
When this condition holds, lifter $i$ optimally declares an attempt weight slightly exceeding the expected $w_H$; otherwise, lifter $i$ declares a weight slightly exceeding $y_H$.

\begin{hypothesis}
    When the higher rival's expected attempt weight exceeds lifter $i$'s current best, an increase in competitive pressure from above---measured by $w_H - y_i$---induces lifter $i$ to choose an attempt weight at least as heavy as $y_H$. Moreover, when the higher rival is expected to succeed with high probability, or when lifter $i$'s own success probability at $y_H$ is low, lifter $i$ is more likely to engage in even greater risk-taking by declaring an attempt weight exceeding the higher rival's expected attempt.
\end{hypothesis}

\subsection{Lifting stage}
We now turn to the lifting stage. In this stage, lifter $i$ chooses execution effort $e_i\ge 0$ to maximize expected utility, taking the declared attempt weight $w_i$ as given. Let $p_i(e_i,w_i)\in[0,1]$ denote lifter $i$'s probability of successfully lifting weight $w_i$. We assume $p_{i,e}(e_i,w_i)>0$ and $p_{i,w}(e_i,w_i)<0$, i.e., higher effort increases the success probability while heavier weights reduce it. Effort incurs a cost $C(e_i)$ with $C'(e_i)>0$ and $C''(e_i)>0$.

As in the attempt stage, interim rank serves as the reference point. The only payoff-relevant outcomes are whether $x-r=1$, $x-r=-1$, or $x-r=0$. The utility consequences of these outcomes are therefore given by the value function $v(\cdot)$ defined above.

We consider two benchmark situations. Under pressure from below (the risk of being overtaken), success prevents a rank loss while failure leads to a rank loss; expected utility is
\begin{align*}
    EU(e_i;w_i)
    = \bigl(1-p_i(e_i,w_i)\bigr)\cdot(-\lambda) - C(e_i).
\end{align*}
Under pressure from above, a rank gain occurs if $w_i>w_H$ and lifter $i$ succeeds, or if $y_H<w_i<w_H$ and lifter $i$ succeeds while the higher rival fails; otherwise the rank remains unchanged. Expected utility is therefore
\begin{align*}
    EU(e_i;w_i)
    = p_i(e_i,w_i)\bigl(\mathbf{1}(w_i>w_H)+\mathbf{1}(y_H<w_i<w_H)(1-q_H(w_H))\bigr)\cdot 1 - C(e_i).
\end{align*}

In either case, the optimal effort $e_i^*$ satisfies the first-order condition
\begin{align*}
p_{i,e}(e_i^*,w_i)\cdot V = C'(e_i^*),
\end{align*}
where $V=\lambda>1$ under pressure from below and $V=\mathbf{1}(w_i>w_H)+\mathbf{1}(y_H<w_i<w_H)(1-q_H(w_H))\le 1$ under pressure from above.

We further assume that the marginal effect of effort on success probability is weakly decreasing, i.e., $p_{i,ee}(e_i,w_i)\le 0$, which captures diminishing returns to effort. Under this standard regularity condition and assuming an interior solution, the implicit function theorem implies that
\begin{align*}
    \frac{\partial e_i^*}{\partial V}
    =
    \frac{p_{i,e}(e_i^*,w_i)}{C''(e_i^*)-V\,p_{i,ee}(e_i^*,w_i)}
    >0.
\end{align*}
Therefore, the possibility of a rank change increases optimal effort relative to a situation in which rank remains unchanged regardless of the outcome (i.e., $V=0$). Moreover, under loss-averse preferences ($\lambda>1$), effort responds more strongly to pressure from below than to pressure from above.

\begin{hypothesis}
At the lifting stage, competitive pressure arising from potential rank changes increases execution effort and, consequently, the probability of success relative to situations in which rank is unaffected by the outcome. Furthermore, under loss-averse preferences, pressure from the lower rival (the risk of being overtaken) leads to a larger increase in success probability than pressure from the higher rival.
\end{hypothesis}

\section{Estimation}

\subsection{Choice of attempt weight}
As the first empirical exercise, we employ a linear regression model to estimate the effect of pressure on the choice of attempt weight. 
We regress the outcome on the observed characteristics as follows: 
\begin{align}
    \tilde{W}_{it}^{k} = X_{it}\beta + Z_{it}^{k}\gamma + \varepsilon_{it}, \label{eq:regression_attempt}
\end{align}
where $\tilde{W}_{it}^{k}$ is the difference between lifter $i$'s attempt weight and personal best as of attempt $k$ in competition $t$, and $X_{it}$ is a vector of observed characteristics of lifter $i$ and competition $t$, including gender, body weight, number of competition experiences, a dummy variable for first participation, and fixed effects for competition $t$'s equipment category, age class, division, weight class, and federation. It also includes a dummy variable indicating whether the lifter shares the same declared attempt weight with another competitor, potentially causing ambiguity in the lifting order due to unresolved tie-breaking rules, and another dummy variable capturing mismatches between the actual lifting order and the interim ranking order, which may affect the set of observable information about rivals. $Z_{it}^{k}$ is a pressure variable during the attempt stage, defined later; $\varepsilon_{it}$ is assumed to be an i.i.d. error; and $\beta$ and $\gamma$ are vectors of coefficients on $X_{it}$ and $Z_{it}^{k}$.
A larger $\tilde{W}_{it}^{k}$ implies that the lifter chooses a more challenging attempt weight.
Our primary interest lies in $\gamma$, that is, the sensitivity to pressure. 

The pressure variables, $Z_{it}^{k}$, are exogenous due to the sequential game setting and are defined as rival-to-self gaps relative to the focal lifter's current best outcome at the decision point. Specifically, pressure from the lower-ranked rival is measured as the lower rival's declared attempt weight minus the focal lifter's current best successful outcome, and pressure from the higher-ranked rival is measured as the expected higher rival attempt weight minus the focal lifter's current best successful outcome. The actual higher-ranked attempt weight should not be used because it is not realized at lifter $i$'s attempt, although it must be predictable. We discuss the predictability issues and how to correct prediction errors, in particular, for constructing standard errors via the bootstrap in Section \ref{sec:robustness}. These variables capture the potential for the lifter's rank to change by overtaking or being overtaken by the rivals.

\subsection{Success probability}

As the second empirical exercise, we use a linear probability model to estimate the probability of successfully lifting the attempt weight. We regress the outcome on the observed characteristics as follows:
\begin{align}
    Y_{it} = X_{it}\beta + \tilde{Z}_{it}^{k}\gamma + \tilde{W}_{it}^{k} \delta + \eta_{it},
\end{align}
where $Y_{it}$ is whether the attempt is successful ($Y_{it}=1$) or not ($Y_{it}=0$), $\tilde{Z}_{it}^{k}$ is a pressure variable during lifting, $\eta_{it}$ is assumed to be an i.i.d. error, and $\delta$ represents the coefficient for the difference between lifter $i$'s attempt weight and his personal best. Note that $\tilde{W}_{it}^{k}$, the choice of attempt weight, may be correlated with unobserved body conditions in $\eta$, potentially causing endogeneity issues. To address this, we employ a two-stage least squares (2SLS) regression, using Equation \eqref{eq:regression_attempt} as the first-stage regression, with the exogenous pressure variables during the attempt, $Z_{it}^{k}$, serving as instrumental variables (IV).
Our IV strategy requires the exclusion restriction that, conditional on $X_{it}$ and $\tilde{Z}_{it}^{k}$, attempt-stage pressure $Z_{it}^{k}$ affects success probability $Y_{it}$ only through the chosen attempt weight $\tilde{W}_{it}^{k}$. This restriction is justified by two features of our design: (1) we control for lifting-stage pressure through $\tilde{Z}_{it}^{k}$, which captures realized rank-change pressure at execution, and (2) we intentionally model attempt and lifting stages separately, so any remaining direct channel from $Z_{it}^{k}$ to $Y_{it}$ is assumed to be limited after conditioning on $X_{it}$ and $\tilde{Z}_{it}^{k}$.
 
The pressure variable $\tilde{Z}_{it}^{k}$ during the lift consists of two indicator variables. The first, $\mathbf{1}(\text{Turned around})_{it}^{k}$, equals one if a lower-ranked rival's $k$-th successful attempt exceeds lifter $i$'s current best outcome, and zero otherwise. When this indicator equals one, the lower rival has already recorded a lift above lifter $i$'s current best at that stage, so lifter $i$ would be overtaken in the interim ranking if he fails to improve upon his current best in the ongoing attempt.
The second, $\mathbf{1}(\text{Turning around})_{it}^{k}$, equals one if lifter $i$'s $k$-th attempt weight exceeds the higher-ranked rival's current best outcome, and zero otherwise. When this indicator equals one, a successful lift would place lifter $i$'s outcome above the higher rival's current best, potentially reversing their ranking depending on the outcome of the higher rival's subsequent attempt. These variables capture the potential for rank changes through being overtaken or overtaking, respectively.

\section{Results}

\subsection{Choice of attempt weight}

\begin{table}[!htbp]
  \begin{center}
      \caption{Regression of the Difference between the Second and Third Attempt Weight Compared to the Personal Best on Pressure}
      \label{tb:estimate_attempt_substituted_bench1kg_list_with_lifter_event_num_experience_attending_competition_dummy_zero_current_best_without_bootstrap} 
      
\begin{tabular}[t]{lcc}
\toprule
  & (1) & (2)\\
\midrule
Dependent Variable & $W_{it}^{2}-$(best) & $W_{it}^{3}-$(best)\\
Male & -1.279*** & 3.583***\\
 & (0.409) & (0.595)\\
Body weight & 0.269*** & 0.373***\\
 & (0.023) & (0.035)\\
Num experience & 0.306*** & 0.325***\\
 & (0.085) & (0.112)\\
1(first participation) & 32.122*** & 33.623***\\
 & (1.507) & (1.633)\\
Pressure, lower rival, 2nd & 0.106*** & \\
 & (0.009) & \\
Pressure, higher rival, 2nd & 0.449*** & \\
 & (0.010) & \\
Pressure, lower rival, 3rd &  & 0.057***\\
 &  & (0.006)\\
Pressure, higher rival, 3rd &  & 0.409***\\
 &  & (0.011)\\
\midrule
Control & X & X\\
Num.Obs. & 246521 & 246521\\
\bottomrule
\end{tabular}

  \end{center}\footnotesize
  \textit{Notes}: We control for gender, body weight, equipment category, age class, division, weight class, federation, and indicators for tie in declared weights and mismatch between lifting order and interim ranking. Standard errors in brackets are clustered at the federation level. $W_{it}^{2}$ and $W_{it}^{3}$ are lifter $i$'s attempt weights in the second and third attempts in competition $t$. Pressure from the lower rival at the second attempt is (lower rival second-attempt weight) - (focal lifter's current best outcome after the first attempt). Pressure from the lower rival at the third attempt is (lower rival third-attempt weight) - (focal lifter's current best outcome after the second attempt). Pressure from the higher rival at the second attempt is (predicted higher rival second-attempt weight) - (focal lifter's current best outcome after the first attempt). Pressure from the higher rival at the third attempt is (predicted higher rival third-attempt weight) - (focal lifter's current best outcome after the second attempt). $^{*}p<0.1$; $^{**}p<0.05$; $^{***}p<0.01$.
\end{table} 

Table \ref{tb:estimate_attempt_substituted_bench1kg_list_with_lifter_event_num_experience_attending_competition_dummy_zero_current_best_without_bootstrap} presents the regression results for the difference between the second and third attempt weights compared to the personal best, incorporating various pressure variables. Male lifters exhibit mixed responses, with a statistically significant negative coefficient for the second attempt (–1.279) and a positive coefficient for the third attempt (3.583). This suggests that male lifters adopt a conservative approach in their second attempt but take greater risks in their third attempt, a pattern consistent with the literature on gender differences in risk-taking. Body weight and experience are positively associated with higher attempt weights in both attempts, indicating that heavier and more experienced lifters generally select heavier weights. The positive coefficients for first participation (32.122 and 33.623) suggest that first-time competitors attempt significantly higher weights relative to their current best.

For the second attempt, pressure from both lower- and higher-ranked rivals has a statistically significant positive effect (0.106 and 0.449, respectively), suggesting that the presence of competitive pressure encourages lifters to attempt heavier weights. In the third attempt, pressure from lower-ranked rivals continues to have a positive effect (0.057), though its magnitude is smaller than in the second attempt. Meanwhile, pressure from higher-ranked rivals remains strongly positive (0.409), reinforcing the idea that lifters attempt heavier weights when their higher-ranked competitors perform well. The persistence of positive coefficients in the third attempt suggests that lifters under strong competitive pressure, particularly from higher-ranked rivals, do not necessarily adopt a conservative strategy but instead continue to take risks, likely in an effort to improve their final ranking.

\subsection{Success probability of lifting the attempt weight}

\begin{table}[!htbp]
  \begin{center}
      \caption{Regression of Success Probability of Second and Third Attempts on Pressure}
      \label{tb:estimate_success_probability_list}
      
\begin{tabular}[t]{lcccc}
\toprule
  & (1) & (2) & (3) & (4)\\
\midrule
Dependent Variable & 1(success 2nd) & 1(success 3rd) & 1(success 2nd) & 1(success 3rd)\\
Male & 0.000 & 0.006 & 0.002 & 0.008\\
 & (0.006) & (0.006) & (0.006) & (0.006)\\
Body weight & -0.001*** & -0.001*** & -0.001*** & -0.001***\\
 & (0.000) & (0.000) & (0.000) & \vphantom{1} (0.000)\\
Num experience & 0.001** & -0.001** & 0.001*** & -0.001*\\
 & (0.000) & (0.000) & (0.000) & (0.000)\\
1(first participation) & -0.004 & -0.005* & 0.014*** & 0.011**\\
 & (0.003) & (0.003) & (0.004) & (0.005)\\
$W_{it}^{2}-$(best) & -0.001*** &  & -0.001*** & \\
 & (0.000) &  & (0.000) & \\
$W_{it}^{3}-$(best) &  & 0.000*** &  & -0.001***\\
 &  & (0.000) &  & (0.000)\\
1(Turned around, 2nd) & 0.014*** &  & 0.013*** & \\
 & (0.003) &  & (0.003) & \\
1(Turning around, 2nd) & -0.012*** &  & -0.005 & \\
 & (0.004) &  & (0.005) & \\
1(Turned around, 3rd) &  & 0.033*** &  & 0.032***\\
 &  & (0.003) &  & (0.003)\\
1(Turning around, 3rd) &  & -0.038*** &  & -0.034***\\
 &  & (0.005) &  & (0.006)\\
\midrule
Control & X & X & X & X\\
IV &  &  & X & X\\
Num.Obs. & 246521 & 246521 & 246521 & 246521\\
R2 & 0.048 & 0.023 & 0.046 & 0.022\\
R2 Adj. & 0.044 & 0.019 & 0.042 & 0.017\\
RMSE & 0.45 & 0.48 & 0.45 & 0.48\\
\bottomrule
\end{tabular}

  \end{center}\footnotesize
  \textit{Notes}: We control for gender, body weight, equipment category, age class, division, weight class, federation, and indicators for tie in declared weights and mismatch between lifting order and interim ranking. Standard errors in brackets are clustered at the federation level. 1(Turned around) equals one if a lower-ranked rival's successful attempt exceeds the lifter's current best outcome, so that the lifter would be overtaken in the ranking if he fails to improve upon his current best in the ongoing attempt. 1(Turning around) equals one if the lifter's attempt weight exceeds the higher-ranked rival's current best outcome, so that a successful lift would place the lifter's best outcome above that of the higher-ranked rival. $^{*}p<0.1$; $^{**}p<0.05$; $^{***}p<0.01$.
\end{table} 

Table \ref{tb:estimate_success_probability_list} provides insights into how pressure influences the success probabilities of the second and third lifting attempts. In Column 1, the pressure from being turned around in the second attempt has a small but positive effect on success probability (0.014), suggesting that lifters may exert additional effort to avoid losing rank. However, the coefficient is modest, indicating a limited influence of loss aversion in this context. Conversely, the pressure of turning around a higher-ranked lifter in the second attempt has a slightly negative coefficient (–0.012), implying that lifters attempting to overtake a higher rival may face increased difficulty in execution.

For the third attempt, the pressure of being turned around remains positively associated with success probability (0.033), while the pressure of turning around a higher-ranked rival has a negative coefficient (–0.038). This suggests that lifters are more likely to succeed when pressured from below but face greater challenges when attempting to surpass a higher-ranked competitor. Columns 3 and 4, which present the results of the IV regression, show consistent patterns with slightly reduced magnitudes and statistical significance.

Overall, these findings highlight the role of pressure across attempts, with patterns broadly consistent with loss-averse preferences and overtaking higher-ranked rivals remaining particularly challenging. The asymmetry—where upward pressure consistently lowers success probabilities,\footnote{\textcolor{black}{While the conceptual framework predicts that pressure from a higher-ranked rival increases effort, increased effort does not necessarily translate into higher success probabilities. In particular, choking under pressure may offset the effort response, resulting in lower observed success rates under upward pressure.}} while downward pressure has neutral or slightly positive effects—is consistent with the loss-averse utility structure assumed in the theoretical framework, whereby individuals exert more effort under pressure from lower-ranked rivals to avoid losses.\footnote{This pattern is also observed in other competitive settings, such as marathons \citep{allen2017reference}.}

\subsection{Heterogeneity}

Tables \ref{tb:estimate_attempt_substituted_bench1kg_list_with_lifter_event_num_experience_attending_competition_dummy_zero_current_best_heterogeneity} and \ref{tb:estimate_success_probability_substituted_bench1kg_list_with_lifter_event_num_experience_attending_competition_dummy_zero_current_best_iv_heterogeneity} show how responses to pressure vary by gender, experience, and historical rivalry frequency. Male lifters respond more strongly to both lower- and higher-ranked rivals (0.104 and 0.455, respectively) than female lifters (0.068 and 0.257), particularly under upward pressure. More experienced lifters are more responsive to higher-ranked rivals, highlighting the role of accumulated competitive cues. A higher historical rivalry frequency—defined as the number of prior encounters with a rival—is associated with greater responsiveness, suggesting that familiarity influences attempt selection. These trends indicate that male, experienced lifters facing well-known rivals take greater risks.

For success probabilities, gender differences vary by pressure direction. When under pressure from lower-ranked rivals—i.e., at risk of being overtaken—male lifters show higher success probabilities in both second (0.015) and third attempts (0.032). Female lifters, by contrast, show little to no benefit in the second attempt (–0.003) but exhibit a positive effect in the third (0.034), suggesting that downward pressure converges across genders in later stages.

In contrast, when attempting to overtake higher-ranked rivals, both genders experience reduced success probabilities, slightly more negative for female lifters (–0.039 vs. –0.034 in the third attempt). This suggests that upward pressure imposes psychological and performance costs for most lifters, with gender gaps persisting, while motivational effects of downward pressure are more stage-dependent and gender-convergent.

Experience plays a limited role in moderating these effects, with small but significant coefficients. Historical rivalry frequency is linked to higher success under downward pressure (0.012 in the second attempt; 0.028 in the third), but lower success under upward pressure (–0.007 in the third). This suggests familiarity with lower-ranked rivals helps manage performance anxiety, whereas familiarity with higher-ranked rivals may heighten psychological stress. Overall, these results underscore the complex interplay between gender, experience, and competitive history in shaping risk-taking and execution under pressure.

\begin{landscape}
{
\begin{table}[!htbp]
  \begin{center}
      \caption{Attempt Heterogeneity on Pressure}
      \label{tb:estimate_attempt_substituted_bench1kg_list_with_lifter_event_num_experience_attending_competition_dummy_zero_current_best_heterogeneity}
      
\begin{tabular}[t]{lcccccc}
\toprule
  & (1) & (2) & (3) & (4) & (5) & (6)\\
\midrule
Dependent Variable & $W_{it}^{2}-$(best) &  &  & $W_{it}^{3}-$(best) &  & \\
(Pressure, lower rival)$\times$ 1(female) & 0.068*** &  &  & -0.021 &  & \\
 & (0.015) &  &  & (0.026) &  & \\
(Pressure, lower rival)$\times$ 1(male) & 0.104*** &  &  & 0.056*** &  & \\
 & (0.009) &  &  & (0.006) &  & \\
(Pressure, higher rival)$\times$ 1(female) & 0.257*** &  &  & 0.154*** &  & \\
 & (0.010) &  &  & (0.015) &  & \\
(Pressure, higher rival)$\times$ 1(male) & 0.455*** &  &  & 0.417*** &  & \\
 & (0.010) &  &  & (0.011) &  & \\
(Pressure, lower rival)$\times$ (Num experience) &  & 0.004*** &  &  & 0.001 & \\
 &  & (0.001) &  &  & (0.001) & \\
(Pressure, higher rival)$\times$ (Num experience) &  & 0.031*** &  &  & 0.032*** & \\
 &  & (0.001) &  &  & (0.002) & \\
(Pressure, lower rival)$\times$ (Historical rivalry freq) &  &  & 0.100*** &  &  & 0.127***\\
 &  &  & (0.011) &  &  & (0.010)\\
(Pressure, higher rival)$\times$ (Historical rivalry freq) &  &  & 0.095*** &  &  & 0.095***\\
 &  &  & (0.012) &  &  & (0.015)\\
\midrule
Control & X & X & X & X & X & X\\
Num.Obs. & 247449 & 247449 & 247449 & 247438 & 247438 & 247438\\
R2 & 0.535 & 0.400 & 0.332 & 0.346 & 0.292 & 0.256\\
R2 Adj. & 0.533 & 0.397 & 0.329 & 0.343 & 0.289 & 0.253\\
\bottomrule
\end{tabular}

  \end{center}\footnotesize
  \textit{Notes}: We control for gender, body weight, equipment category, age class, division, weight class, federation, and indicators for tie in declared weights and mismatch between lifting order and interim ranking as in the main specification. Standard errors in brackets are clustered at the federation level. $^{*}p<0.1$; $^{**}p<0.05$; $^{***}p<0.01$.
\end{table} 
}
\end{landscape}

\begin{landscape}
{
\begin{table}[!htbp]
  \begin{center}
      \caption{Success Probability Heterogeneity on Pressure}
      \label{tb:estimate_success_probability_substituted_bench1kg_list_with_lifter_event_num_experience_attending_competition_dummy_zero_current_best_iv_heterogeneity}
      
\begin{tabular}[t]{lcccccc}
\toprule
  & (1) & (2) & (3) & (4) & (5) & (6)\\
\midrule
Dependent Variable & 1(success 2nd) &  &  & 1(success 3rd) &  & \\
1(Turned around)$\times$ 1(female) & -0.003 &  &  & 0.034*** &  & \\
 & (0.006) &  &  & (0.013) &  & \\
1(Turned around)$\times$ 1(male) & 0.015*** &  &  & 0.032*** &  & \\
 & (0.003) &  &  & (0.004) &  & \\
1(Turning around)$\times$ 1(female) & -0.014** &  &  & -0.039*** &  & \\
 & (0.007) &  &  & (0.010) &  & \\
1(Turning around)$\times$ 1(male) & -0.003 &  &  & -0.034*** &  & \\
 & (0.006) &  &  & (0.006) &  & \\
1(Turned around)$\times$ (Num experience) &  & 0.001*** &  &  & 0.003*** & \\
 &  & (0.000) &  &  & (0.001) & \\
1(Turning around)$\times$ (Num experience) &  & 0.001* &  &  & -0.002*** & \\
 &  & (0.000) &  &  & (0.000) & \\
1(Turned around)$\times$ (Historical rivalry freq) &  &  & 0.012*** &  &  & 0.028***\\
 &  &  & (0.003) &  &  & (0.007)\\
1(Turning around)$\times$ (Historical rivalry freq) &  &  & -0.002 &  &  & -0.007\\
 &  &  & (0.004) &  &  & (0.004)\\
\midrule
Control & X & X & X & X & X & X\\
IV & X & X & X & X & X & X\\
Num.Obs. & 246521 & 246521 & 246521 & 246521 & 246521 & 246521\\
R2 & 0.046 & 0.046 & 0.046 & 0.022 & 0.021 & 0.021\\
R2 Adj. & 0.042 & 0.041 & 0.041 & 0.017 & 0.016 & 0.016\\
\bottomrule
\end{tabular}

  \end{center}\footnotesize
  \textit{Notes}: We control for gender, body weight, equipment category, age class, division, weight class, federation, and indicators for tie in declared weights and mismatch between lifting order and interim ranking as in the main specification. Standard errors in brackets are clustered at the federation level. $^{*}p<0.1$; $^{**}p<0.05$; $^{***}p<0.01$.
\end{table} 

}
\end{landscape}

\section{Counterfactual}

Individual responses to pressure influence the overall distribution of risk-taking, success probability, and expected outcomes. The unique structure of official bench press competitions provides an opportunity to disentangle these effects and compare alternative competition settings to the current format.

Using the estimated results, we first simulate attempt weights and success probabilities for each attempt of each lifter under actual pressure.\footnote{Our regression results use the predicted attempt weight minus the personal best as a dependent variable. We obtain simulated attempt weight by adding the personal best to the predicted outcomes.} We then calculate expected achieved weights, defined as the product of the predicted attempt weight and the success probability. These serve as benchmark outcomes under actual pressure conditions.

The first alternative competition setting assumes that lifters would have to simultaneously submit their planned attempt weights for all attempts before the first attempt and commit to the plan. This setting eliminates the possibility of pressure from rivals influencing weight attempts. While introducing this as a formal rule is unrealistic, such a commitment is common at the individual level. We refer to this scenario as the ``no pressure during attempt'' setting.

The second alternative competition setting posits that lifters could not observe the attempt weights and outcomes of their rivals after the competition begins. This removes the influence of sequentially updated outcomes from rivals on success probability. Similar to the first scenario, formalizing this rule would be challenging, though such a commitment is feasible at the individual level. We refer to this scenario as the ``no pressure during lifting'' setting.

The third alternative competition setting combines the first two scenarios, creating a condition in which there would be no pressure during both the attempt and the lifting phases.

For these alternative competition settings, we simulate attempt weights, success probabilities, and expected achieved weights for each lifter, holding the estimated coefficients on the pressure variables at zero.

\subsection{Attempt and success probability}

\begin{figure}[!ht]
\begin{center}
  \subfloat[Attempt]{\includegraphics[height = 0.28\textheight]{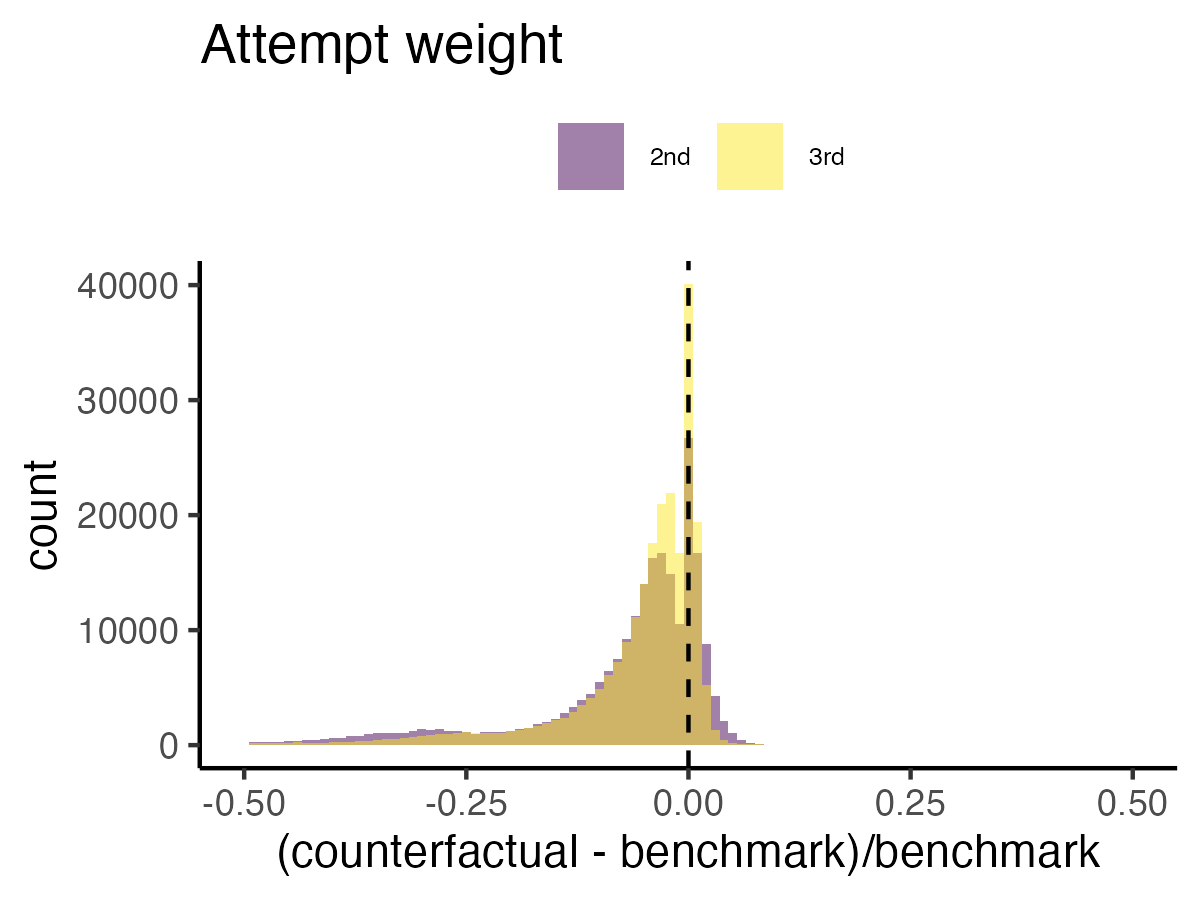}}
  \subfloat[Lifting]{\includegraphics[height = 0.28\textheight]{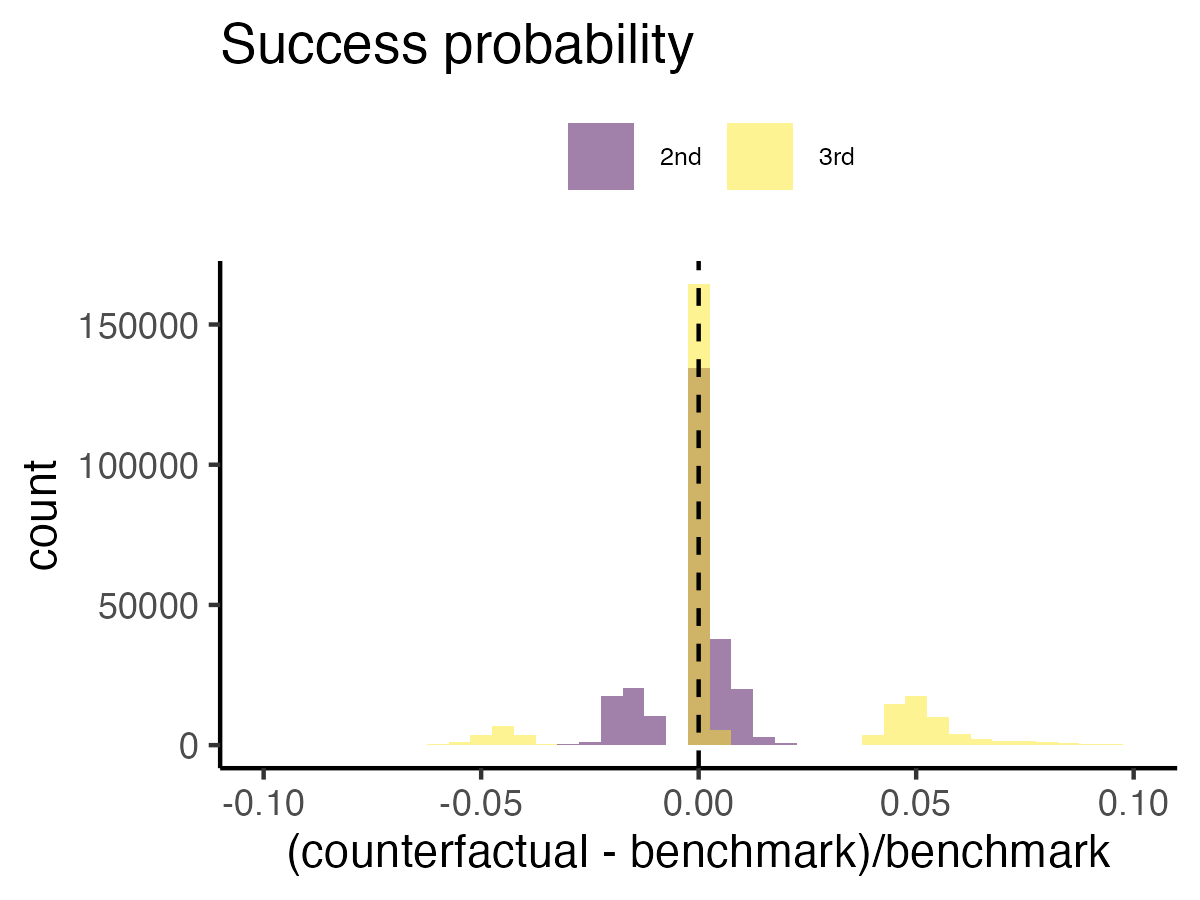}}
\end{center}
\caption{Counterfactual Attempt and Lifting}\footnotesize
\textit{Notes}: We use estimated coefficients in Table \ref{tb:estimate_attempt_substituted_bench1kg_list_with_lifter_event_num_experience_attending_competition_dummy_zero_current_best_without_bootstrap} and Columns (3) and (4) in Table \ref{tb:estimate_success_probability_list}.
\label{fg:change_rate_diff_lifter_event_apply_bench2kg_minus_current_benchmaxkg_fitted}
\end{figure}

Panel (a) in Figure \ref{fg:change_rate_diff_lifter_event_apply_bench2kg_minus_current_benchmaxkg_fitted} shows the distribution of proportional changes in second and third attempt weights under the counterfactual without pressure during the attempt. Both distributions exhibit substantial deviations from zero, indicating that pressure has a strong influence on attempt weights. The third attempt distribution is slightly more concentrated around zero, but the overall shapes of the second and third attempt distributions are similar. These patterns suggest that pressure during the attempt significantly affects weight selection, with comparable impacts on second and third attempts.

Panel (b) in Figure \ref{fg:change_rate_diff_lifter_event_apply_bench2kg_minus_current_benchmaxkg_fitted} presents the distribution of proportional changes in the second and third attempt success probabilities from the benchmark to the counterfactual. For the second attempt, the distribution is relatively concentrated, with peaks around zero, indicating that some lifters experience modest gains or losses in success probability when pressure is removed. In contrast, the third attempt shows a more dispersed distribution with heavier tails, suggesting greater variability in how lifters respond to the absence of pressure. While a subset of lifters benefit, many experience reduced success. Overall, the removal of competitive pressure leads to mixed effects in both attempts, with more pronounced heterogeneity in the third.

\subsection{Expected weight}

\begin{figure}[!ht]
\begin{center}
\includegraphics[height = 0.28\textheight]{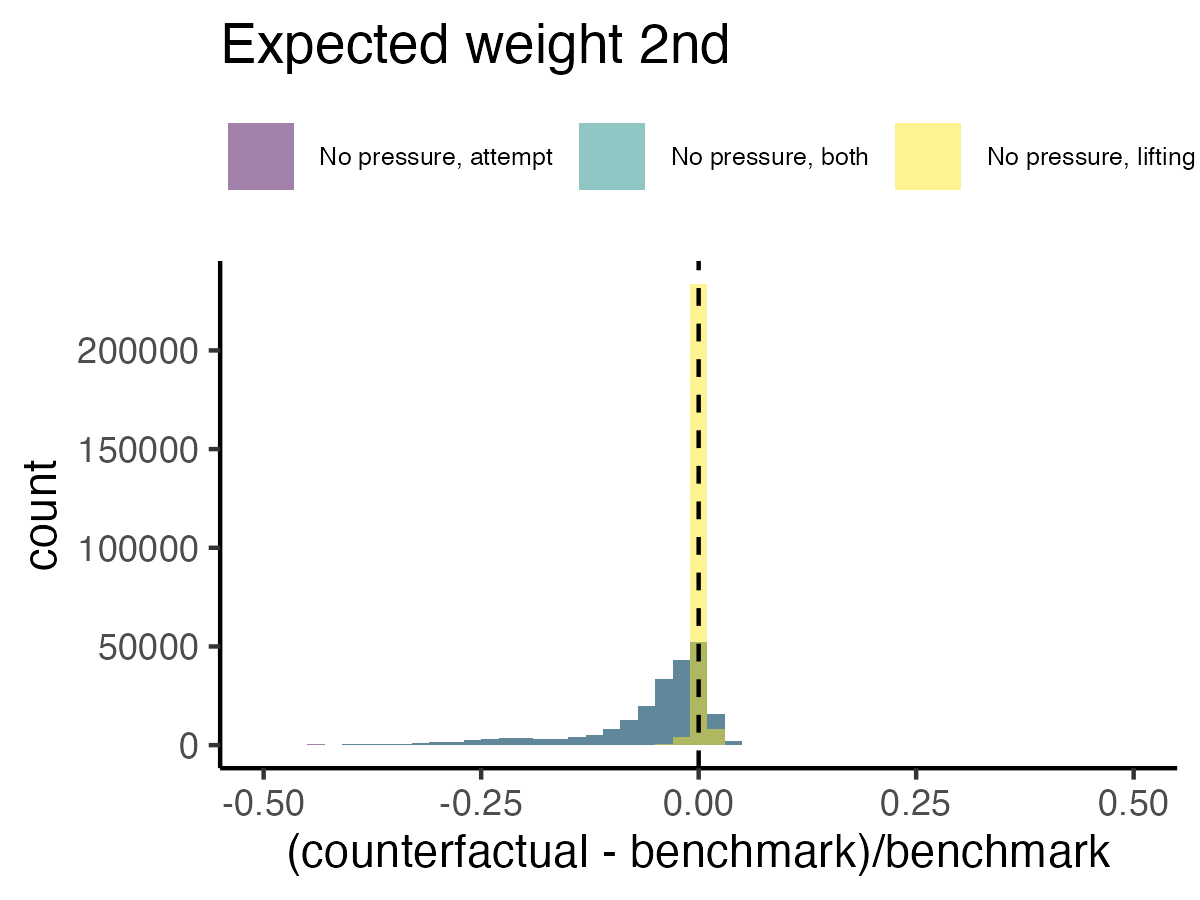}
\includegraphics[height = 0.28\textheight]{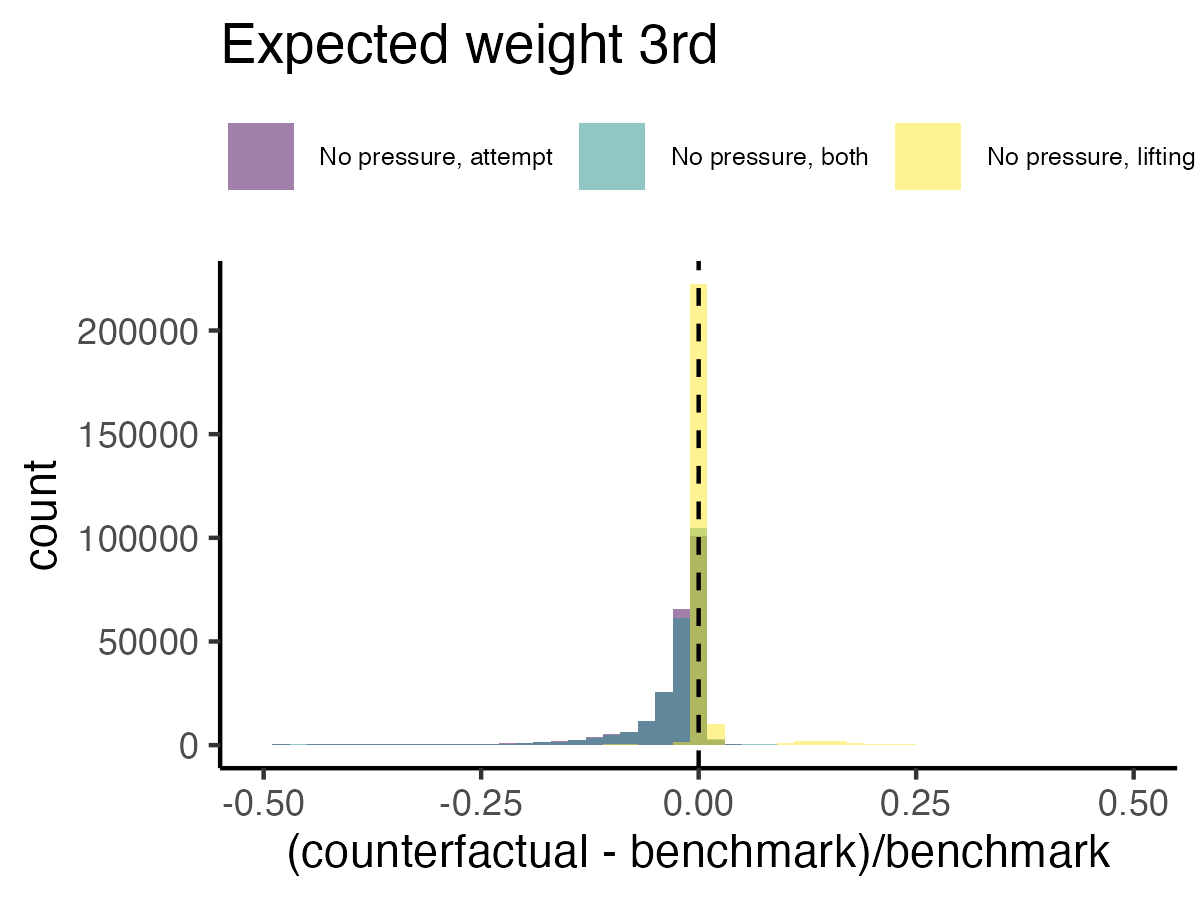}
\end{center}
\caption{Counterfactual Expected Achieved Weight}\footnotesize
\textit{Notes}: We use estimated coefficients in Table \ref{tb:estimate_attempt_substituted_bench1kg_list_with_lifter_event_num_experience_attending_competition_dummy_zero_current_best_without_bootstrap} and Columns (3) and (4) in Table \ref{tb:estimate_success_probability_list}. Positive values indicate better performance under no pressure than under actual competition pressure.
\label{fg:change_rate_expected_diff_lifter_event_apply_bench2kg_minus_current_benchmaxkg_no_pressure_attempt}
\end{figure}

Figure \ref{fg:change_rate_expected_diff_lifter_event_apply_bench2kg_minus_current_benchmaxkg_no_pressure_attempt} presents the distribution of expected achieved weight differences for second and third attempts, comparing the conditions of ``no pressure'' during the attempt stage, during the lifting stage, or during both stages. 

In the second attempt, most counterfactual outcomes are negative, especially in the ``no pressure during both stages'' condition, where the distribution is clearly left-skewed. The ``no pressure during lifting'' case shows a sharper peak near zero, with only limited gains on the positive side.

In the third attempt, the negative side remains dominant, particularly under ``no pressure during attempt'' and ``no pressure during both stages'' conditions. The ``no pressure during lifting'' case still includes some positive outcomes, but most observations are at or below zero. These patterns highlight substantial heterogeneity in pressure responses and reinforce the motivational role of real-time competition.

As practical advice, most lifters may benefit from adjusting their attempts in response to competitive pressure, as removing pressure tends to reduce both attempt weights and expected achieved weights. However, reacting to pressure from rivals is not universally optimal across lifters. More broadly, individuals in sequential high-stakes settings should account for their own sensitivity to competitive pressure, recognizing that for many, performance deteriorates in its absence.

\section{Robustness check}\label{sec:robustness}

\subsection{Alternative personal best definition}
Our main specification uses each lifter's all-time personal best as an ability-based baseline for measuring deviations in attempt weight. However, older records may not accurately reflect a lifter's current ability, so we check robustness by replacing the all-time personal best used in the dependent variable with the trailing twelve-month personal best---the highest outcome achieved within the 12 months preceding the competition. The pressure variables remain unchanged, as they are defined in terms of rivals' declared or predicted attempt weights and the focal lifter's realized best outcome within the competition.

\begin{table}[!htbp]
  \begin{center}
      \caption{Regression of the Difference between Attempt Weight and Trailing Twelve-Month Personal Best on Pressure}
      \label{tb:estimate_attempt_substituted_bench1kg_list_with_lifter_event_num_experience_attending_competition_dummy_zero_current_best_without_bootstrap_one_year_best} 
      
\begin{tabular}[t]{lcc}
\toprule
  & (1) & (2)\\
\midrule
Dependent Variable & $W_{it}^{2}-$(best) & $W_{it}^{3}-$(best)\\
Male & 0.528 & 5.785***\\
 & (0.324) & (0.476)\\
Body weight & 0.151*** & 0.254***\\
 & (0.031) & (0.035)\\
Num experience & -0.059 & -0.125*\\
 & (0.057) & (0.067)\\
1(first participation) & 29.646*** & 30.688***\\
 & (1.287) & (1.344)\\
Pressure, lower rival, 2nd & 0.092*** & \\
 & (0.009) & \\
Pressure, higher rival, 2nd & 0.483*** & \\
 & (0.008) & \\
Pressure, lower rival, 3rd &  & 0.043***\\
 &  & (0.005)\\
Pressure, higher rival, 3rd &  & 0.427***\\
 &  & (0.010)\\
\midrule
Control & X & X\\
Num.Obs. & 230744 & 230744\\
R2 & 0.561 & 0.325\\
R2 Adj. & 0.559 & 0.322\\
\bottomrule
\end{tabular}

  \end{center}\footnotesize
  \textit{Notes}: We control for gender, body weight, equipment category, age class, division, weight class, federation, and indicators for tie in declared weights and mismatch between lifting order and interim ranking. Standard errors in brackets are clustered at the federation level. $W_{it}^{2}$ and $W_{it}^{3}$ are lifter $i$'s attempt weights in the second and third attempts in competition $t$. Pressure from the lower rival at the second attempt is (lower rival second-attempt weight) - (focal lifter's current best outcome after the first attempt). Pressure from the lower rival at the third attempt is (lower rival third-attempt weight) - (focal lifter's current best outcome after the second attempt). Pressure from the higher rival at the second attempt is (predicted higher rival second-attempt weight) - (focal lifter's current best outcome after the first attempt). Pressure from the higher rival at the third attempt is (predicted higher rival third-attempt weight) - (focal lifter's current best outcome after the second attempt). $^{*}p<0.1$; $^{**}p<0.05$; $^{***}p<0.01$.
\end{table}

\begin{table}[!htbp]
  \begin{center}
      \caption{Regression of Success Probability of Second and Third Attempts on Pressure, Trailing Twelve-Month Personal Best}
      \label{tb:estimate_success_probability_list_one_year_best}
      
\begin{tabular}[t]{lcccc}
\toprule
  & (1) & (2) & (3) & (4)\\
\midrule
Dependent Variable & 1(success 2nd) & 1(success 3rd) & 1(success 2nd) & 1(success 3rd)\\
Male & 0.003 & 0.008 & 0.005 & 0.010\\
 & (0.007) & (0.006) & (0.007) & (0.007)\\
Body weight & -0.001*** & -0.001*** & -0.001*** & -0.001***\\
 & (0.000) & (0.000) & (0.000) & \vphantom{1} (0.000)\\
Num experience & 0.002*** & 0.000 & 0.002*** & 0.000\\
 & (0.000) & (0.000) & (0.000) & (0.000)\\
1(first participation) & 0.012*** & 0.010*** & 0.021*** & 0.019***\\
 & (0.003) & (0.003) & (0.003) & (0.005)\\
$W_{it}^{2}-$(best) & -0.001*** &  & -0.001*** & \\
 & (0.000) &  & (0.000) & \\
$W_{it}^{3}-$(best) &  & 0.000*** &  & -0.001***\\
 &  & (0.000) &  & (0.000)\\
1(Turned around, 2nd) & 0.012*** &  & 0.012*** & \\
 & (0.003) &  & (0.003) & \\
1(Turning around, 2nd) & -0.007* &  & -0.003 & \\
 & (0.004) &  & (0.005) & \\
1(Turned around, 3rd) &  & 0.033*** &  & 0.032***\\
 &  & (0.004) &  & (0.004)\\
1(Turning around, 3rd) &  & -0.036*** &  & -0.033***\\
 &  & (0.005) &  & (0.006)\\
\midrule
Control & X & X & X & X\\
IV &  &  & X & X\\
Num.Obs. & 230744 & 230744 & 230744 & 230744\\
R2 & 0.043 & 0.022 & 0.042 & 0.021\\
R2 Adj. & 0.038 & 0.017 & 0.037 & 0.016\\
\bottomrule
\end{tabular}

  \end{center}\footnotesize
  \textit{Notes}: We control for gender, body weight, equipment category, age class, division, weight class, federation, and indicators for tie in declared weights and mismatch between lifting order and interim ranking. Standard errors in brackets are clustered at the federation level. $^{*}p<0.1$; $^{**}p<0.05$; $^{***}p<0.01$.
\end{table} 

Table \ref{tb:estimate_attempt_substituted_bench1kg_list_with_lifter_event_num_experience_attending_competition_dummy_zero_current_best_without_bootstrap_one_year_best} corresponds to Table \ref{tb:estimate_attempt_substituted_bench1kg_list_with_lifter_event_num_experience_attending_competition_dummy_zero_current_best_without_bootstrap} for the analysis of attempt weights using the trailing twelve-month personal best. The main findings remain robust: pressure from both lower- and higher-ranked rivals continues to have positive and significant effects on attempt weights in both the second and third attempts. While the coefficient on the male dummy changes sign in the second attempt, the overall patterns are consistent, reinforcing that lifters respond to competitive pressure, particularly from higher-ranked rivals.

Similarly, Table \ref{tb:estimate_success_probability_list_one_year_best} corresponds to Table \ref{tb:estimate_success_probability_list} for the analysis of success probabilities, utilizing the trailing twelve-month personal best. The main results are largely consistent with the primary specification. Upward pressure—especially attempts to overtake higher-ranked rivals—is associated with a higher likelihood of failure.

Overall, the alternative specification using the trailing twelve-month personal best confirms the key findings from the main analysis: competitive pressure affects both strategic weight selection and execution success, and these effects persist across time horizons.

\subsection{Expected rivals' attempt weights}\label{sec:expected_rival_attempt_weight}
In our empirical exercises, we use the predicted attempt weight of lifter $i$'s higher-ranked rival—unobservable to lifter $i$ at the attempt stage. Although this approach involves prediction errors, assuming that the higher rival's attempt weight is completely unknown would ignore the pressure exerted by the higher-ranked rival, which plays a significant role in actual competition.

\begin{table}[!htbp]
  \begin{center}
      \caption{Prediction of Attempt Weights Using Linear Regression}
      \label{tb:prediction_of_attempt_weight} 
      
\begin{tabular}{llrr}
\toprule
Attempt & Model & RMSE & R2\\
\midrule
1st & Full & 23.837 & 0.813\\
2nd & Full & 4.214 & 0.994\\
2nd & Without 1st success & 4.481 & 0.994\\
3rd & Full & 3.450 & 0.996\\
3rd & Without 2nd success & 4.191 & 0.995\\
\bottomrule
\end{tabular}

  \end{center}\footnotesize
  \textit{Notes}: All models include gender, body weight, equipment category, age class, division, weight class, federation, a dummy for first participation, number of competition experiences, and personal best as predictors. They also include a dummy variable indicating whether the lifter shares the same declared attempt weight with another competitor---potentially causing ambiguity in the lifting order due to unresolved tie-breaking rules---and another dummy variable capturing mismatches between the actual lifting order and the interim ranking order, which may affect the set of observable information about rivals. In the full model for the 2nd attempt, both the attempt weight and the success of the 1st attempt are included as predictors, while in the ``without 1st success'' model, the success of the 1st attempt is excluded. Similarly, in the full model for the 3rd attempt, the attempt weight and the success of the 2nd attempt are used, while in the ``without 2nd success'' model, the success of the 2nd attempt is excluded. Predictive accuracy is assessed using 5-fold cross-validation.
\end{table} 

Table \ref{tb:prediction_of_attempt_weight} presents the Root Mean Squared Error (RMSE) and $R^2$ values for different prediction models of attempt weights using linear regression. Predictive accuracy is assessed through 5-fold cross-validation. All models include gender, body weight, equipment category, age class, division, weight class, federation, a dummy for first participation, number of competition experiences, and personal best as predictors. They also include a dummy variable indicating whether the lifter shares the same declared attempt weight with another competitor---potentially causing ambiguity in the lifting order due to unresolved tie-breaking rules---and another dummy variable capturing mismatches between the actual lifting order and the interim ranking order, which may affect the set of observable information about rivals. In the full model for the 2nd attempt, both the attempt weight and the success of the 1st attempt are included as predictors, while in the ``without 1st success'' model, the success of the 1st attempt---unobservable when predicting the higher rival's 2nd attempt weight---is excluded. Similarly, in the full model for the 3rd attempt, the attempt weight and the success of both the 1st and 2nd attempts are used, whereas in the ``without 2nd success'' model, the success of the 2nd attempt---unobservable when predicting the higher rival's 3rd attempt weight---is excluded.

The prediction for the 1st attempt yields an RMSE of 23.837 and an $R^2$ of 0.813, indicating that accurately predicting the 1st attempt remains challenging. In contrast, the predictions for the 2nd and 3rd attempts achieve an $R^2$ exceeding 0.99 across all models, demonstrating that even without incorporating the outcome of the preceding attempt, which is unobservable when predicting the higher rival's attempt weight, high predictive accuracy can still be maintained. Comparing the full models with the ``without'' models for both the 2nd and 3rd attempts, the full models exhibit slightly lower RMSEs (4.214 vs. 4.481 for the 2nd attempt; 3.450 vs. 4.191 for the 3rd attempt), indicating marginally better predictive accuracy. This suggests that the outcome of an attempt reflects the lifter's condition on the day of the competition, influencing weight selection, while lower-ranked rivals are unable to observe this condition. These findings support our assumption that the pressure variable---derived from predictions of the higher rival's attempt weight---is independent of unobserved factors such as the lifter's condition and can be considered exogenous.

\begin{table}[!htbp]
  \begin{center}
      \caption{Regression of the Difference between the Second and Third Attempt Weight Compared to the Personal Best on Pressure with Bootstrap}
      \label{tb:estimate_attempt_substituted_bench1kg_list_with_lifter_event_num_experience_attending_competition_dummy_zero_current_best} 
      
\begin{tabular}[t]{lcc}
\toprule
  & (1) & (2)\\
\midrule
Dependent Variable & $W_{it}^{2}-$(best) & $W_{it}^{3}-$(best)\\
Male & -1.477*** & 3.210***\\
 & (0.384) & (0.529)\\
Body weight & 0.274*** & 0.355***\\
 & (0.014) & (0.033)\\
Num experience & 0.229*** & 0.255***\\
 & (0.052) & (0.098)\\
1(first participation) & 32.849*** & 34.061***\\
 & (0.989) & (0.984)\\
Pressure, lower rival, 2nd & 0.111*** & \\
 & (0.007) & \\
Pressure, higher rival, 2nd & 0.440*** & \\
 & (0.009) & \\
Pressure, lower rival, 3rd &  & 0.068***\\
 &  & (0.007)\\
Pressure, higher rival, 3rd &  & 0.407***\\
 &  & (0.010)\\
\midrule
Control & X & X\\
Num.Obs. & 246521 & 246521\\
R2 & 0.536 & 0.356\\
R2 Adj. & 0.534 & 0.353\\
\bottomrule
\end{tabular}

  \end{center}\footnotesize
  \textit{Notes}: We control for gender, body weight, equipment category, age class, division, weight class, federation, and indicators for tie in declared weights and mismatch between lifting order and interim ranking. $ W_{it}^{2} $ and $ W_{it}^{3} $ are lifter $ i $'s attempt weights in the second and third attempts in competition $ t $. Pressure from the lower rival at the second attempt is (lower rival second-attempt weight) - (focal lifter's current best outcome after the first attempt). Pressure from the lower rival at the third attempt is (lower rival third-attempt weight) - (focal lifter's current best outcome after the second attempt). Pressure from the higher rival at the second attempt is (predicted higher rival second-attempt weight) - (focal lifter's current best outcome after the first attempt). Pressure from the higher rival at the third attempt is (predicted higher rival third-attempt weight) - (focal lifter's current best outcome after the second attempt). For higher rivals, predicted attempt weights are obtained as the mean from the bootstrap procedure, while standard errors are computed as the standard deviation of the bootstrap estimates, ensuring that the uncertainty in the prediction process is fully reflected in the inference. $^{*}p<0.1$; $^{**}p<0.05$; $^{***}p<0.01$.
\end{table} 

To address the prediction error problem of using predicted values as explanatory variables, we implement a bootstrap procedure when incorporating the predicted attempt weight of lifter $ i $'s higher-ranked rival---unobservable to lifter $ i $ at the $ k $-th attempt stage---into our regression framework. Since this predicted value, denoted by $ \hat{Z}_{it}^{k} $, is derived from a separate estimation and contains prediction error, standard errors tend to be downward biased due to the smoothing effect of prediction. To correct for this, we apply a bootstrap approach that resamples the original data at the federation level, re-estimates the prediction model for $ \hat{Z}_{it}^{k} $ in each bootstrap iteration, and then regresses the lifter's own attempt weight or success probability on the re-estimated $ \hat{Z}_{it}^{k} $. This procedure ensures that the uncertainty in $ \hat{Z}_{it}^{k} $ is incorporated into second-stage inference by reflecting its sampling variability.

Compared to Table \ref{tb:estimate_attempt_substituted_bench1kg_list_with_lifter_event_num_experience_attending_competition_dummy_zero_current_best_without_bootstrap}, which uses predicted higher rivals' attempt weights, Table \ref{tb:estimate_attempt_substituted_bench1kg_list_with_lifter_event_num_experience_attending_competition_dummy_zero_current_best} incorporates a bootstrap procedure to account for uncertainty in predicted higher rival attempt weights. Standard errors are broadly similar in magnitude (some coefficients are larger and others are smaller), while point estimates and signs remain consistent. Notably, the male coefficient in the second attempt shifts from –1.279 to –1.477, and pressure from higher-ranked rivals in the second attempt is stable, reinforcing the robustness of competitive pressure effects. Similarly, the effect of pressure from higher-ranked rivals in the third attempt remains stable (0.407 vs. 0.409 in Table \ref{tb:estimate_attempt_substituted_bench1kg_list_with_lifter_event_num_experience_attending_competition_dummy_zero_current_best_without_bootstrap}). The stability of estimates confirms the validity of the benchmark results even when accounting for prediction uncertainty.

\subsection{Alternative definitions of risk-taking}

We also examine whether our results are robust to an alternative scaling of the dependent variable. Instead of the level difference $W_{it}^{k} - \text{(best)}$, we rescale this difference by the within-weight-class standard deviation of personal bests, yielding $(W_{it}^{k} - \text{(best)}) / \text{sd}(\text{best})$. \textcolor{black}{Because the dispersion of personal bests differs across weight classes, the same absolute deviation may reflect a different degree of risk-taking in classes with substantially different ability levels. In our regression specification, weight-class fixed effects absorb systematic level differences across classes, while this rescaling adjusts for differences in the dispersion of personal bests within classes.} Tables \ref{tb:estimate_attempt_robustness_depvar_2nd} and \ref{tb:estimate_attempt_robustness_depvar_3rd} report the results. Column (1) reproduces the baseline specification and Column (2) uses the standardized difference. The coefficients on pressure from both lower- and higher-ranked rivals remain positive and statistically significant across both specifications and for both attempts. The $R^2$ values are comparable between the two specifications, and the qualitative conclusions are unchanged: competitive pressure significantly increases attempt weight selection under both scalings.

\begin{table}[!htbp]
  \begin{center}
      \caption{Robustness: Alternative Dependent Variable, Second Attempt}
      \label{tb:estimate_attempt_robustness_depvar_2nd}
      
\begin{tabular}[t]{lcc}
\toprule
  & (1) & (2)\\
\midrule
Dependent Variable & $W_{it}^{2}-$(best) & $(W_{it}^{2}-$(best))/$\text{sd}$(best)\\
Male & -1.254*** & 0.012\\
 & (0.460) & (0.012)\\
Body weight & 0.267*** & 0.003***\\
 & (0.025) & (0.000)\\
Num experience & 0.166 & 0.001\\
 & (0.115) & (0.002)\\
1(first participation) & 31.479*** & 0.547***\\
 & (1.372) & (0.025)\\
Pressure, lower rival & 0.101*** & 0.001***\\
 & (0.008) & (0.000)\\
Pressure, higher rival & 0.425*** & 0.007***\\
 & (0.012) & (0.000)\\
\midrule
Control & X & X\\
Num.Obs. & 204215 & 204217\\
R2 & 0.514 & 0.495\\
R2 Adj. & 0.512 & 0.493\\
\bottomrule
\end{tabular}

  \end{center}\footnotesize
  \textit{Notes}: Column (1) uses the level difference $W_{it}^{2} - \text{(best)}$ as the dependent variable. Column (2) standardizes the level difference by the within-weight-class standard deviation of personal bests: $(W_{it}^{2} - \text{(best)}) / \text{sd}(\text{best})$. We control for gender, body weight, equipment category, age class, division, weight class, federation, and indicators for tie in declared weights and mismatch between lifting order and interim ranking. $^{*}p<0.1$; $^{**}p<0.05$; $^{***}p<0.01$.
\end{table}

\begin{table}[!htbp]
  \begin{center}
      \caption{Robustness: Alternative Dependent Variable, Third Attempt}
      \label{tb:estimate_attempt_robustness_depvar_3rd}
      
\begin{tabular}[t]{lcc}
\toprule
  & (1) & (2)\\
\midrule
Dependent Variable & $W_{it}^{3}-$(best) & $(W_{it}^{3}-$(best))/$\text{sd}$(best)\\
Male & 3.439*** & 0.103***\\
 & (0.633) & (0.014)\\
Body weight & 0.347*** & 0.005***\\
 & (0.049) & (0.001)\\
Num experience & 0.178 & 0.001\\
 & (0.149) & (0.002)\\
1(first participation) & 35.362*** & 0.608***\\
 & (1.821) & (0.032)\\
Pressure, lower rival & 0.071*** & 0.001***\\
 & (0.008) & (0.000)\\
Pressure, higher rival & 0.415*** & 0.007***\\
 & (0.012) & (0.000)\\
\midrule
Control & X & X\\
Num.Obs. & 181232 & 181233\\
R2 & 0.369 & 0.352\\
R2 Adj. & 0.366 & 0.350\\
\bottomrule
\end{tabular}

  \end{center}\footnotesize
  \textit{Notes}: Column (1) uses the level difference $W_{it}^{3} - \text{(best)}$ as the dependent variable. Column (2) standardizes the level difference by the within-weight-class standard deviation of personal bests: $(W_{it}^{3} - \text{(best)}) / \text{sd}(\text{best})$. We control for gender, body weight, equipment category, age class, division, weight class, federation, and indicators for tie in declared weights and mismatch between lifting order and interim ranking. $^{*}p<0.1$; $^{**}p<0.05$; $^{***}p<0.01$.
\end{table}

\section{Conclusion}

Using data from bench press competitions, our study highlights the impact of competitive pressure on lifters' attempt selections and success probabilities. Pressure from both lower- and higher-ranked rivals in the second attempt encourages more aggressive weight selections. In the third attempt, lifters continue to take risks under pressure from both directions, with higher-ranked rivals in particular inducing significant upward adjustments in attempt weights.

Heterogeneity analysis reveals that male lifters, experienced competitors, and those with repeated rivalry encounters respond more strongly to pressure. Male lifters take greater risks, while female and less experienced lifters remain more conservative. Familiarity with rivals amplifies responsiveness to competition. While success probabilities tend to decline under upward pressure for all lifters, the effect is notably stronger among women, underscoring the psychological difficulty of overtaking stronger competitors.

Counterfactual simulations show that without competitive pressure, lifters tend to adopt more conservative attempts. However, the effects on performance are highly heterogeneous: while some lifters benefit from the absence of pressure, others experience lower success rates and reduced expected outcomes. These results highlight that external competition plays a dual role—motivating some lifters to perform better while destabilizing others—underscoring the importance of individual sensitivity in shaping performance under pressure.

Future research could explore structural estimation of decision-making across attempts. Lifters strategically allocate their stamina, a key aspect in SBD (Squat-Bench Press-Deadlift) competitions. Modeling this sequential structure—where lifters choose attempt weights, treat each lift as a lottery, and adjust under stamina constraints—could offer deeper insights into risk-taking and performance optimization.


\bibliographystyle{ecca}
\bibliography{weightlifting}

\newpage

\appendix

\section{Additional Results}

\subsection{Alternative definitions of competitive pressure}

As a further robustness check, we consider alternative definitions of the pressure variables. In our baseline specification, pressure from a lower-ranked rival is measured as the difference between the lower rival's declared attempt weight and the focal lifter's current best outcome, while pressure from a higher-ranked rival uses the predicted attempt weight of the higher rival. Here, we replace these with two alternatives: (i) using the rival's personal best record in place of the rival's declared or predicted attempt weight, and \textcolor{black}{(ii) using the rival's best realized outcome in attempts preceding the focal attempt.} Column (1) in Tables \ref{tb:estimate_attempt_robustness_pressure_2nd} and \ref{tb:estimate_attempt_robustness_pressure_3rd} re-estimates the baseline specification on the harmonized sample used in Columns (2) and (3), where rival personal-best and rival best-realized-outcome measures are available. Column (2) uses the rival's personal best, and Column (3) uses the rival's best realized outcome.

\begin{table}[!htbp]
  \begin{center}
      \caption{Robustness: Alternative Pressure Definitions, Second Attempt}
      \label{tb:estimate_attempt_robustness_pressure_2nd}
      
\begin{tabular}[t]{lccc}
\toprule
  & (1) & (2) & (3)\\
\midrule
Dependent Variable & $W_{it}^{2}-$(best) & $W_{it}^{2}-$(best) & $W_{it}^{2}-$(best)\\
Male & -1.254*** & 6.860*** & 3.129***\\
 & (0.460) & (1.222) & (1.042)\\
Body weight & 0.267*** & 0.299*** & 0.539***\\
 & (0.025) & (0.060) & (0.049)\\
Num experience & 0.166 & 0.057 & 0.176\\
 & (0.115) & (0.169) & (0.144)\\
1(first participation) & 31.479*** & 36.194*** & 33.560***\\
 & (1.372) & (1.719) & (1.705)\\
Pressure, lower rival (original) & 0.101*** &  & \\
 & (0.008) &  & \\
Pressure, higher rival (original) & 0.425*** &  & \\
 & (0.012) &  & \\
Pressure, lower rival (rival PB) &  & 0.098*** & \\
 &  & (0.006) & \\
Pressure, higher rival (rival PB) &  & 0.127*** & \\
 &  & (0.004) & \\
Pressure, lower rival (rival best outcome) &  &  & 0.125***\\
 &  &  & (0.014)\\
Pressure, higher rival (rival best outcome) &  &  & 0.323***\\
 &  &  & (0.013)\\
\midrule
Control & X & X & X\\
Num.Obs. & 204215 & 204242 & 204242\\
R2 & 0.514 & 0.370 & 0.375\\
R2 Adj. & 0.512 & 0.368 & 0.373\\
\bottomrule
\end{tabular}

  \end{center}\footnotesize
  \textit{Notes}: We control for gender, body weight, equipment category, age class, division, weight class, federation, and indicators for tie in declared weights and mismatch between lifting order and interim ranking. Column (1) uses the baseline pressure definition, where pressure from a lower-ranked rival is the difference between the lower rival's declared attempt weight and the focal lifter's current best outcome, and pressure from a higher-ranked rival is the difference between the higher rival's predicted attempt weight and the focal lifter's current best outcome. Column (2) replaces the rival's declared or predicted attempt weight with the rival's personal best. Column (3) replaces it with the rival's best realized outcome in attempts preceding the focal attempt. Column (1) is estimated on the same restricted sample as Columns (2) and (3) for comparability, which yields fewer observations than in the full baseline table because rival personal-best information must be available. $^{*}p<0.1$; $^{**}p<0.05$; $^{***}p<0.01$.
\end{table}

\begin{table}[!htbp]
  \begin{center}
      \caption{Robustness: Alternative Pressure Definitions, Third Attempt}
      \label{tb:estimate_attempt_robustness_pressure_3rd}
      
\begin{tabular}[t]{lccc}
\toprule
  & (1) & (2) & (3)\\
\midrule
Dependent Variable & $W_{it}^{3}-$(best) & $W_{it}^{3}-$(best) & $W_{it}^{3}-$(best)\\
Male & 3.439*** & 8.984*** & 5.175***\\
 & (0.633) & (1.339) & (1.023)\\
Body weight & 0.347*** & 0.316*** & 0.500***\\
 & (0.049) & (0.063) & (0.064)\\
Num experience & 0.178 & 0.069 & 0.171\\
 & (0.149) & (0.192) & (0.164)\\
1(first participation) & 35.362*** & 39.355*** & 36.833***\\
 & (1.821) & (2.046) & (1.999)\\
Pressure, lower rival (original) & 0.071*** &  & \\
 & (0.008) &  & \\
Pressure, higher rival (original) & 0.415*** &  & \\
 & (0.012) &  & \\
Pressure, lower rival (rival PB) &  & 0.056*** & \\
 &  & (0.006) & \\
Pressure, higher rival (rival PB) &  & 0.090*** & \\
 &  & (0.005) & \\
Pressure, lower rival (rival best outcome) &  &  & 0.033***\\
 &  &  & \vphantom{1} (0.012)\\
Pressure, higher rival (rival best outcome) &  &  & 0.381***\\
 &  &  & (0.012)\\
\midrule
Control & X & X & X\\
Num.Obs. & 181232 & 181232 & 181232\\
R2 & 0.369 & 0.273 & 0.307\\
R2 Adj. & 0.366 & 0.270 & 0.304\\
\bottomrule
\end{tabular}

  \end{center}\footnotesize
  \textit{Notes}: We control for gender, body weight, equipment category, age class, division, weight class, federation, and indicators for tie in declared weights and mismatch between lifting order and interim ranking. Column (1) uses the baseline pressure definition, where pressure from a lower-ranked rival is the difference between the lower rival's declared attempt weight and the focal lifter's current best outcome, and pressure from a higher-ranked rival is the difference between the higher rival's predicted attempt weight and the focal lifter's current best outcome. Column (2) replaces the rival's declared or predicted attempt weight with the rival's personal best. Column (3) replaces it with the rival's best realized outcome in attempts preceding the focal attempt. Column (1) is estimated on the same restricted sample as Columns (2) and (3) for comparability, which yields fewer observations than in the full baseline table because rival personal-best information must be available. $^{*}p<0.1$; $^{**}p<0.05$; $^{***}p<0.01$.
\end{table}

Across all specifications, the pressure coefficients remain positive and statistically significant at the 1\% level for both the second and third attempts. The magnitudes vary somewhat across definitions, reflecting differences in the information content of each measure. The baseline specification, which uses declared or predicted attempt weights, yields the highest $R^2$, consistent with the fact that actual strategic choices incorporate more competition-relevant information than historical records alone. Nevertheless, the qualitative conclusions are unchanged: competitive pressure from both lower- and higher-ranked rivals significantly increases attempt weight selection.

\subsection{First attempt weight and rival personal best}\label{sec:first_attempt_rival_pb}

We examine whether first-attempt weight selection is associated with the personal bests of adjacent rivals. In our main specification, competitive pressure is defined with respect to potential changes in the interim ranking, which serves as the reference point in the sequential competition structure. Because no interim ranking exists prior to the first attempt, our baseline pressure variables cannot be defined for this stage. For this appendix exercise, we fix the order using declared first-attempt weights (with the official lifting order), define adjacent lower and higher rivals within that fixed order, and then construct pressure variables from personal-best gaps relative to those rivals. Unlike our main specification, rivals' personal bests are determined prior to the competition and do not arise from the sequential realization of outcomes during the competition. As a result, they do not capture the exogenous variation generated by the sequential competition structure, and the results should therefore be interpreted as descriptive rather than causal.

Specifically, pressure from a lower rival is defined as (adjacent lower rival's personal best) - (focal lifter's personal best), and pressure from a higher rival is defined analogously. Table \ref{tb:estimate_attempt_first_attempt_placebo} reports the results. The dependent variable is $W_{it}^{1} - \text{(best)}$, the difference between the first-attempt weight and the lifter's personal best. The coefficients on the rival personal best pressure variables are $0.111$ (lower rival) and $0.071$ (higher rival), both statistically significant at the 1\% level. The positive signs are consistent with the pattern observed in our main second- and third-attempt results: lifters who face stronger rivals---as measured by the gap in personal bests---tend to select higher attempt weights. However, because personal bests are pre-determined and not generated by the within-competition information revelation mechanism that our identification strategy exploits, this finding should be interpreted as a descriptive association rather than a causal effect.

\begin{table}[!htbp]
  \begin{center}
      \caption{First Attempt Weight and Rival Personal Best}
      \label{tb:estimate_attempt_first_attempt_placebo}
      
\begin{tabular}[t]{lc}
\toprule
  & (1)\\
\midrule
Dependent Variable & $W_{it}^{1}-$(best)\\
Male & 22.713***\\
 & (0.974)\\
Body weight & 0.746***\\
 & (0.057)\\
Num experience & 1.090***\\
 & (0.114)\\
1(first participation) & 128.328***\\
 & (3.854)\\
Pressure, lower rival (PB) & 0.111***\\
 & (0.009)\\
Pressure, higher rival (PB) & 0.071***\\
 & (0.007)\\
\midrule
Control & X\\
Num.Obs. & 267342\\
R2 & 0.856\\
R2 Adj. & 0.855\\
\bottomrule
\end{tabular}

  \end{center}\footnotesize
  \textit{Notes}: The dependent variable is the difference between the first-attempt weight and the lifter's personal best ($W_{it}^{1} - \text{(best)}$). Rivals are defined as adjacent competitors in the fixed order based on declared first-attempt weights (using official lifting order when ties occur). Pressure from a lower-ranked rival is the difference between the lower rival's personal best and the focal lifter's personal best; pressure from a higher-ranked rival is defined analogously. We control for gender, body weight, equipment category, age class, division, weight class, federation, number of competition experiences, and an indicator for first-time participation. $^{*}p<0.1$; $^{**}p<0.05$; $^{***}p<0.01$.
\end{table}

\end{document}